\documentstyle[psfig]{mn}                       

\title[The near-IR Ca\,{\normalsize \it II} triplet: atmospheric parameters]
{Empirical calibration of the near-IR Ca\,{\large \bf II} triplet -- II. 
The stellar atmospheric parameters}

\author[A.J. Cenarro et al.]  
    {A.J.~Cenarro,$^1$\thanks{E-mail: cen@astrax.fis.ucm.es} J.~Gorgas,$^1$
    N.~Cardiel,$^1$ S.~Pedraz,$^{1,2}$ R.F.~Peletier,$^3$
\newauthor and A.~Vazdekis.$^4$\\ $^1$Dept. de Astrof\'{\i}sica, Fac. de Ciencias F\'{\i}sicas, Universidad
  Complutense de Madrid, 28040 Madrid, Spain\\ $^2$Calar Alto Observatory,
  CAHA, Apdo. 511, 04004 Almer\'{\i}a, Spain\\ $^3$School of Physics and
  Astronomy, University of Nottingham, University Park, Nottingham NG7 2RD,
  UK\\ $^4$Dept. of Physics, University of Durham, South Road, Durham DH1 3LE,
  UK\\}
\date{}
\pagerange{\pageref{firstpage}--\pageref{lastpage}}
\pubyear{}

\def\LaTeX{L\kern-.36em\raise.3ex\hbox{a}\kern-.15em
    T\kern-.1667em\lower.7ex\hbox{E}\kern-.125emX}

\begin{document}

\label{firstpage}

\maketitle

\begin{abstract}

We present an homogeneous set of stellar atmospheric parameters ({\it
T}$_{\rm eff}$, log\,{\it g}, [Fe/H]) for a sample of about 700 field
and cluster stars which constitute a new stellar library in the
near-infrared developed for stellar population synthesis in this
spectral region ($\lambda$ 8350--9020 \AA). Having compiled the
available atmospheric data in the literature for field stars, we have
found systematic deviations between the atmospheric parameters from
different bibliographic references. The Soubiran, Katz \& Cayrel
(1998) sample of stars with very well determined fundamental
parameters has been taken as our standard reference system, and other
papers have been calibrated and bootstrapped against it. The obtained
transformations are provided in this paper. Once most of the datasets
were on the same system, final parameters were derived by performing
error weighted means.  Atmospheric parameters for cluster stars have
also been revised and updated according to recent metallicity scales
and colour-temperature relations.
\end{abstract}


\section{Introduction}

This paper is the second one in a series devoted to advance in the
understanding of the stellar population properties of composite
stellar systems by studying the strength of the integrated Ca triplet
in the near-infrared spectral range. As we have already explained in
Paper~I (Cenarro et al. 2001a), the main objectives of the series are
to derive empirical fitting functions describing the behaviour of the
Ca triplet index in terms of the stellar atmospheric parameters
(Cenarro et al. 2001b, Paper~III), and to perform stellar populations
synthesis modeling in the near-IR spectral range (Vazdekis et
al. 2001, Paper~IV). An ample stellar library covering a wide range of
atmospheric parameters is necessary to obtain accurate fitting
functions (Worthey et al. 1994, hereafter W94; Gorgas et al. 1999) and
to derive reliable synthetic spectra for stellar populations of
different ages and metallicities (Vazdekis 1999).  Even so, this is
not enough to ensure the quality of the empirical predictions, since
they also depend on the accuracy of the input atmospheric parameters.

Recently, Gorgas et al. \shortcite{Gorgas2} have shown how important
an accurate, homogeneous set of input atmospheric parameters is when
deriving empirical fitting functions.  When calculating fitting
functions for the $\lambda4000$~\AA\ break, and after a thorough
treatment of the errors, these authors show that the residuals from
the derived functions are considerably larger than those just expected
from measurement errors, indicating that the uncertainties in the
input atmospheric parameters are the main source of random errors.
Moreover, it is worth noting that, in order to obtain accurate fitting
functions, not only an homogeneous but also a reliable set of
atmospheric parameters is needed. Although an homogeneous set of
parameters optimizes the scatter of the derived fitting functions, it
does not guarantee that the zero points of the adopted scales are
absent of systematic errors. Therefore it is also important to choose
an absolute scale as reliable as possible.

Up to date, most of the previous authors which use stellar libraries
to model the composite spectra of external galaxies (e.g. D\'{\i}az,
Terlevich \& Terlevich 1989; Gorgas et al. 1993, hereafter G93; W94;
Jones 1997) have hardly tackled the already known problem of
uncertainties in the atmospheric parameters and their implications on
the final predictions of fitting functions. Instead of doing this, the
usual approach has been to choose the parameters of the sample stars
from the most recent bibliographic sources or take the average values,
without checking whether they were on a completely homogeneous
system. As an example, it is common practice to use straight means
from previous parameter compilations (like the one of Cayrel de
Strobel et al. 1997), even though the individual analyses do not
necessarily all have the same quality or are mutually independent. We
refer the reader to the work of Soubiran, Katz and Cayrel (1998)
(hereafter SKC) for a thorough discussion of these and related
problems. Furthermore, systematic deviations among different
bibliographic sources may exist due to the different approaches for
measuring atmospheric parameters.

In this paper we have derived an homogeneous set of stellar
atmospheric parameters for the stellar library presented in
Paper~I. Section~2 introduces the working method and the atmospheric
data compilation for the field stars. In Section\,3 we present a
calibration of the different bibliographic sources. The new
atmospheric parameters for these stars are derived in Section\,4,
whereas in Section\,5 we estimate the uncertainties in the final
parameters. In addition, we have also revised the atmospheric
parameters for the cluster stars in the library (Section\,6). Finally,
Section\,7 is reserved to discussion and summary.

Along this paper, the fundamental atmospheric parameters are
considered, that is, effective temperature ({\it T}$_{\rm eff}$ in K),
surface gravity (log\,{\it g} with {\it g} in cm\,s$^{-2}$) and
metallicity ([Fe/H] = log(Fe/H) -- log(Fe/H)$_{\sun}$).

\section{The method}

The main goal of this paper is to obtain a new and homogeneous set of
atmospheric parameters for the stars of the library introduced in
Paper~I. For this purpose, we have made a compilation of the three
main atmospheric parameters of these stars in the literature. We
selected one article of data, the reference source, which contained a
large number of stars with parameters of high quality and bootstrapped
all other sources against it, to end up with an homogeneous system of
stellar atmospheric parameters. After this, the relative quality of
all other data sources was determined by computing the
r.m.s. deviation from the reference sample. Weighted according to the
data quality, the various data sources were averaged to provide a
final homogeneous set of atmospheric measurements.

The compilation includes 356 bibliographic sources, although not all
of them were finally used to derive the final parameters. To start
with, we included all the data from the catalogue of [Fe/H]
determinations of Cayrel de Strobel et al. (1997), which contains
parameters for more than 3000 stars from 700 sources up to 1995. After
that, since not all our stars were included in the above catalogue,
and to take into account more recent papers, we enlarged the
compilation with 47 additional sources. Unavoidably, we could not
include sources that were published during or after the last steps of
this work (i.e. mid 1999). It must be noted that, even for stars with
data in Cayrel de Strobel et al. (1997), we went back to the original
data sources to exclude references that simply quote previous
determinations.

To calculate systematic deviations, we had to select one standard
source as a reference system. It was essential that this standard
source contained a large number of stars including the three
atmospheric parameters in an homogeneous way, and with a reasonably
large parameter coverage. It is worthwhile to remark the latter since
it is well known that a generic atmospheric parameter can not be
derived independently from the other two ones. Concerning the choice
of a reference system, it is important to keep in mind that our final
purpose is to obtain an empirical calibration of the behaviour of
several line-strength indices as a function of atmospheric parameters,
We are basically interested in ensuring that stars with very similar
spectra have the same atmospheric parameters. This is the main reason
why we have selected the paper of SKC as the initial standard
source. It provides self-consistent atmospheric parameters for a total
of 211 echelle spectra of cool stars (4000~K $<T_{\rm eff}<$ 6300~K)
covering a wide range in gravity and metallicity. Making use of a
reference spectral library (which includes stars with well-known
atmospheric parameters) and input parameters for the target stars
(weighted means of previous determinations from the literature), they
followed an iterative method that takes into account spectral features
comparisons, deriving revised values of effective temperature, gravity
and metallicity for the sample of target stars. See full details of
the above method in Katz et al. (1998). The final atmospheric
parameters are, in the mean, consistent with the literature, and
constitute an homogeneous set in the sense that similar spectra have
similar parameters and the other way round.

Therefore, the atmospheric parameters from SKC will be our initial
reference system. To obtain statistically significant comparisons
between SKC and other sources, not only the 108 stars from SKC in
common with our stellar library were included but also the rest of the
stars in their catalogue. Obviously, these calibrations will only be
valid for stars in the effective temperature range spanned by the
sample of SKC, i.e. from 4000~K to 6300~K.  We did not follow a fully
automatic approach and the original parameters for every star were
checked for inconsistencies or outliers, removing original references
when necessary.

\section{Calibration of the different sources}
\label{calibracion}

Once the compilation was finished, we selected for each of the three
atmospheric parameters those references that had at least 25 stars in
common with the complete sample of SKC. That minimum number was chosen
to ensure that comparisons between any source and SKC were
statistically significant.

Let $p$ and $p_{\rm ref}$ be generic atmospheric parameters from any
literature and the reference system (SKC in this first iteration)
respectively. In order to calibrate this source onto the reference
system we determined the following two types of fits and their
significance level, $\alpha$:

\begin{enumerate}
\item A linear fit, $p = A + B\,p_{\rm ref}$.
\item An offset, $p = A + p_{\rm ref}$.
\end{enumerate}

We then tested whether the slope $B$ was significantly different from
1, using a $t$-test and a significance level of $\alpha = 0.1$. If
that was the case, we adopted relation (i) to bootstrap the data from
the source against the reference system ($p^{\ast}=(p - A)/B$, where
$p^{\ast}$ is the corrected parameter). If it was not significant, we
used the same procedure to test the significance of the offset term
$A$ in relation (ii) and applied that correction if necessary
($p^{\ast}=p - A$). Obviously, the original parameters were kept when
this term was not statistically different from 0 ($p^{\ast}=p$).

The procedure detailed above leads to a set of corrected parameters
which were used to calculate the final parameters of a large number of
stars not included in SKC. These stars constitute a new reference set
of stars (hereafter RF1), with parameters in the same system as SKC.
In order to calibrate all those reference sources which did not have
enough stars in common with SKC, the whole process was repeated using
SKC and RF1 together as the reference samples. Since the number of
stars in the remaining sources is generally rather small, the minimum
number of stars in common required to calibrate a reference was
decreased to 15. In this way, we derived a second set of final
parameters which is called RF2. We did not perform further iterations
since, after the second one, those sources that had not been
calibrated yet did not possess enough stars in common with the
reference systems (SKC, RF1 and RF2) to ensure reliable calibrations.

\begin{figure*}                 
\centerline{
\psfig{figure=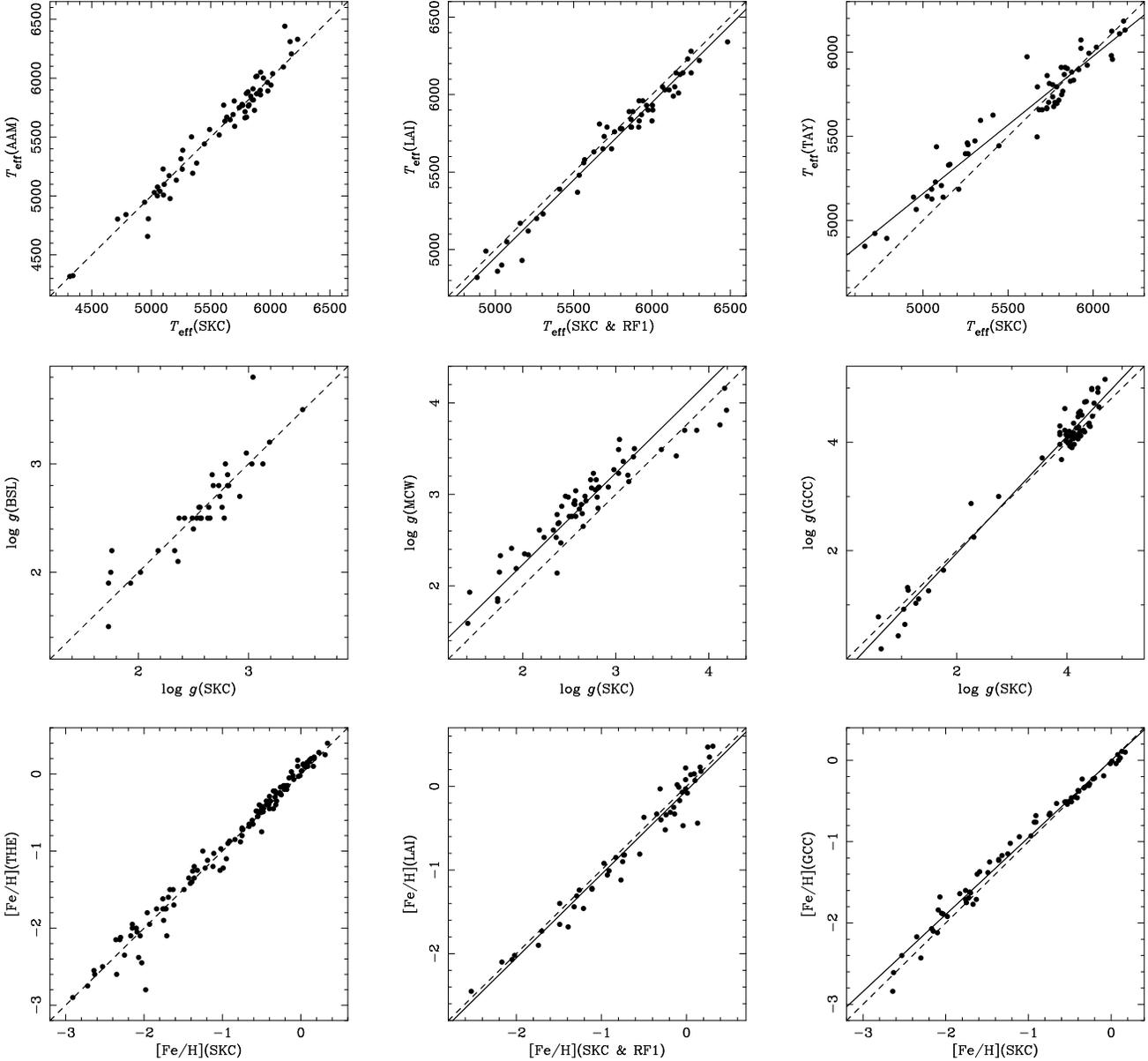}
}
\caption{Calibrations onto the reference system (SKC or SKC \& RF1). The upper
three panels show representative cases of null, offset and linear
corrections for sources giving effective temperatures. A dashed line
show the expected behaviour when no systematic deviation exists,
whereas the solid line is the applied correction. The same is done for
sources publishing surface gravities and metallicities in the centred
and lower panels respectively. Points deviating more than 3$\sigma$
from an initial fit were not used to derive the final
calibration. More details about these and other calibrations are
presented in Tables~\ref{caltemp}, \ref{calgrav} and \ref{calmetal}.}
\label{compref}
\end{figure*}

In Tables~\ref{caltemp}, \ref{calgrav} and \ref{calmetal} we present,
repectively, the details of the calibrations on effective temperature,
gravity and metallicity for all the calibrated sources. Reference
codes are explained in Table~\ref{calrefer}. To illustrate the
procedure, in Fig.~\ref{compref}, we show some representative
calibrations for each atmospheric parameter and kind of correction
that was applied. In the above tables we include a code to indicate
the different methods used to derive the original atmospheric
parameters in each paper. Note that, although the tabulated standard
deviations are due to uncertainties both in the SKC parameters and in
the calibrated reference, a relative comparison of the different
values can provide an estimate of the reliability of the different
methods. Even though a critical analysis of these techniques is out of
the scope of this paper, it must be noted that we do not find any
systematic trend when comparing the uncertainties ($\sigma$) or the
calibration parameters ($A$, $B$) of the different working methods.

\begin{scriptsize}                                                                                       
\begin{table}                                                                                            
\centering{ 
\caption{Calibrations of bibliographic sources to convert their effective
temperatures onto the reference system. Columns are: Reference code
(see Table~\ref{calrefer}), method used to derive temperatures,
number of stars in common with the standard source, applied correction
(s: straight line; o: offset; n: none), standard source (1: SKC; 2:
SKC \& RF1), r.m.s. standard deviation from the fit, independent term,
slope, and range of the fit. Codes for the methods: (a) Infrared
flux method, (b) spectroscopic methods, and (c) from colour relations.
Values from JON and WOR only
include original determinations, that is, parameters taken from other
sources were not employed (this also holds for Tables~2 and~3).}
\label{caltemp}
\begin{tabular}{@{}l@{}c@{}r@{}c@{}crrlc@{}}
\hline
 Code & M & N  &\ \ \ Fit&\ \ \ \ S& {\large $\sigma$}\ \ &$A$\,\ &\ $B$& {\it T}$_{\rm eff}$ \\
\hline                
AAM   &\ a\ \ \ &  67 &\ \ \ \ n\ \ &\ \ \ \ 1& 98.&  0.0  &  1.0 &4300 , 6400\\                        
AFG   &\ b\ \ \ &  30 &\ \ \ \ n\ \ &\ \ \ \ 1&124.&  0.0  &  1.0 &5600 , 6400\\                        
BAL   &\ c\ \ \ &  21 &\ \ \ \ n\ \ &\ \ \ \ 2&100.&  0.0  &1.0   &6000 , 6400\\
BLL   &\ a\ \ \ &  44 &\ \ \ \ s\ \ &\ \ \ \ 2& 75.&--175.5&1.0440&3900 , 6400\\
BSL   &\ c\ \ \ &  39 &\ \ \ \ s\ \ &\ \ \ \ 1& 66.& 396.5 &0.9118&4000 , 5100\\                        
CLL   &\ c\ \ \ &  40 &\ \ \ \ n\ \ &\ \ \ \ 1& 76.&  0.0  &  1.0 &4600 , 6300\\                        
EAG   &\ c\ \ \ &  36 &\ \ \ \ o\ \ &\ \ \ \ 1& 60.& 39.9  & 1.0  &5650 , 6350\\                        
GCC   &\ c\ \ \ &  65 &\ \ \ \ s\ \ &\ \ \ \ 1& 86.&--178.8&1.0397&4100 , 6500\\                        
GRJ   &\ b\ \ \ &  28 &\ \ \ \ s\ \ &\ \ \ \ 2&115.&835.8  &0.8637&5100 , 6300\\
GRS   &\ c\ \ \ &  25 &\ \ \ \ n\ \ &\ \ \ \ 1&116.&  0.0  & 1.0  &3800 , 6100\\                        
HEA   &\ c\ \ \ &  26 &\ \ \ \ s\ \ &\ \ \ \ 2& 65.& 811.0 &0.8529&5100 , 6200\\
JON   &\ c\ \ \ &  47 &\ \ \ \ s\ \ &\ \ \ \ 2& 67.&--291.4&1.0604&4200 , 5300\\
LAI   &\ c\ \ \ &  53 &\ \ \ \ o\ \ &\ \ \ \ 2& 71.&--51.1 &  1.0 &4700 , 6400\\
LCH   &\ c\ \ \ &  38 &\ \ \ \ o\ \ &\ \ \ \ 2& 62.&--72.9 &1.0   &3900 , 5000\\
MAS   &\ c\ \ \ &  38 &\ \ \ \ s\ \ &\ \ \ \ 1& 83.& 2852.0&0.5450&5900 , 6300\\
MCW   &\ c\ \ \ &  62 &\ \ \ \ n\ \ &\ \ \ \ 1& 86.&  0.0  &  1.0 &3900 , 5900\\
NHS   &\ c\ \ \ &  22 &\ \ \ \ n\ \ &\ \ \ \ 2& 96.&  0.0  &1.0   &4700 , 6300\\
PET   &\ c\ \ \ &  26 &\ \ \ \ o\ \ &\ \ \ \ 2&106.&--83.7 &1.0   &4500 , 6400\\
PSB   &\ c\ \ \ &  26 &\ \ \ \ s\ \ &\ \ \ \ 1&101.&  517.7&0.9042&4300 , 6000\\
PSK  &\ bc\ \ \ &  26 &\ \ \ \ s\ \ &\ \ \ \ 2&100.&--404.6&1.0910&4200 , 5300\\
RMB  &\ bc\ \ \ &  25 &\ \ \ \ o\ \ &\ \ \ \ 2& 80.&--77.1 & 1.0  &5200 , 6150\\
SIC   &\ b\ \ \ &  20 &\ \ \ \ s\ \ &\ \ \ \ 2&112.&--661.1&1.1006&4200 , 6300\\
TAY &\ abc\ \ \ &  62 &\ \ \ \ s\ \ &\ \ \ \ 1& 92.& 1075.9&0.8166&4800 , 6200\\
TLL   &\ c\ \ \ &  22 &\ \ \ \ o\ \ &\ \ \ \ 2& 82.&--67.5 &1.0   &4700 , 6300\\
WOR   &\ c\ \ \ &  44 &\ \ \ \ n\ \ &\ \ \ \ 1& 74.&  0.0  &  1.0 &4100 , 6100\\
\hline                                          
\end{tabular}                                                                                      
}
\end{table}
\end{scriptsize}

\begin{scriptsize}                                                                                       
\begin{table}                                                                                            
\centering{ 
\caption{Calibrations of bibliographic sources to convert their surface gravities
onto the reference system. Columns are the same as in
Table~\ref{caltemp}. Methods employed to derive gravities: (a)
spectroscopic method, (b) physical method (parallaxes), (c) physical
method (luminosities from photometric indices), (d) physical method
(luminosities from Ca K line), (e) photometric, and (f) other.}
\label{calgrav}
\begin{tabular}{@{}lcrcccr@{}llc@{}}
\hline
 Code & M & N &Fit&S & {\large $\sigma$}&\multicolumn{2}{c}{$A$}&\ $B$& log\,{\it g}\\
\hline
AFG & a  &  30 &n&1  &  0.27  &  0.&0  &  1.0 &   2.5 ,  4.8    \\                                                
BAL & c  &  23 &n&2  &   0.07 &  2.&560&0.391 &   3.9  ,  4.3   \\          
BSL & cd &  39 &n&1  &  0.19  &  0.&0  &  1.0 &   1.4 ,  3.9    \\                                                
EAG & f  &  36 &o&1  &  0.12  &  0.&042& 1.0  &   3.9 ,  4.6    \\                                                
GCC & a  &  65 &s&1  &  0.24  &--0.&200& 1.077&   0.0 ,  5.2    \\                                                
GRS & b  &  24 &o&1  &  0.30  &  0.&139& 1.0  &   0.7 ,  4.5    \\                                                
HEA & b  &  23 &n&2  &   0.18 &  0.&0  &1.0   &   3.8 ,   4.6   \\          
KNK & e  &  28 &o&1  &  0.14  &  0.&075& 1.0  &   4.0 ,  4.7    \\                                                
LAI & ab &  48 &s&2  &   0.32 &  2.&038& 0.520&   3.4  ,  4.6   \\          
LBO & a  &  16 &n&2  &   0.39 &  0.&0  &1.0   &   0.0  ,  4.0   \\                                
LCH & ad &  38 &o&2  &   0.37 &--0.&527&1.0   &   0.2  ,  2.8   \\
MAS & e  &  38 &o&1  &  0.40  &  0.&247& 1.0  &   3.8 ,  5.0    \\
MCW & bd &  62 &o&1  &  0.21  &  0.&233&  1.0 &   1.6 ,  4.2    \\
NHS & b  &  18 &s&2  &   0.25 &  1.&740&0.609 &   3.7 ,  4.7   \\
PSK & cf &  25 &n&2  &   0.25 &  0.&0  &1.0   &   0.2 , 3.0  \\
TLL & a  &  22 &s&2  &   0.21 &--0.&910&1.210 &   2.5  ,  5.1   \\
WOR & f  &  34 &n&1  &  0.33  &  0.&0  &  1.0 &   1.0 ,  4.8    \\
\hline                                          
\end{tabular}                                                                                      
}
\end{table}
\end{scriptsize}

\begin{scriptsize}                                                                                       
\begin{table}                                                                                            
\centering{ 
\caption{Calibrations of bibliographic sources to convert their metallicities 
onto the reference system. Columns are the same as in
Table~\ref{caltemp}. Methods employed to compute metallicities:
(a) high resolution ($< 0.5$~\AA) spectroscopy, (b) mid resolution
($> 0.5$~\AA) spectroscopy, (c) photometry, and (d) spectrophotometry.}
\label{calmetal}
\begin{tabular}{@{}l@{}c@{}r@{}c@{}ccr@{}llr@{}r@{}}
\hline
 Code & M & N  &\ \ \ Fit&\ \ \ \ \ S & {\large $\sigma$}& &$A$&\ \ $B$& \multicolumn{2}{c}{\ \ [Fe/H]}\\
\hline
AAM &\ \ ac\ \ \ &\  68&\ \ \ \ s\ \ &\ \ \ \ 1&  0.22  &--0.&006& 1.065&--3.0 ,&\ +0.4   \\  
AFG &\ \  a\ \ \ &\  30&\ \ \ \ s\ \ &\ \ \ \ 1&  0.13  &--0.&120& 0.858&--2.5 ,&\ --0.4  \\  
BAL &\ \  a\ \ \ &\  23&\ \ \ \ o\ \ &\ \ \ \ 2&   0.10 &--0.&067&1.0   &--0.7 ,&\ +0.3   \\  
BKP &\ \  b\ \ \ &\  27&\ \ \ \ s\ \ &\ \ \ \ 1&  0.21  &--0.&324& 0.829&--3.1 ,&\ --1.0  \\  
BSL &\ \  a\ \ \ &\  39&\ \ \ \ n\ \ &\ \ \ \ 1&  0.19  &  0.&0  &  1.0 &--0.8 ,&\ +0.5   \\  
CGC &\ \  a\ \ \ &\  27&\ \ \ \ o\ \ &\ \ \ \ 2&   0.10 &  0.&129&1.0   &--2.4 ,&\ --1.0  \\  
CLL &\ \  a\ \ \ &\  41&\ \ \ \ s\ \ &\ \ \ \ 1&  0.10  &  0.&029& 1.070&--2.7 ,&\ +0.2   \\  
EAG &\ \  a\ \ \ &\  36&\ \ \ \ s\ \ &\ \ \ \ 1&  0.05  &--0.&047& 0.925&--1.1 ,&\ +0.2   \\  
GCC &\ \  a\ \ \ &\  65&\ \ \ \ s\ \ &\ \ \ \ 1&  0.10  &--0.&002& 0.947&--3.0 ,&\ +0.2   \\  
GRS &\ \  a\ \ \ &\  25&\ \ \ \ n\ \ &\ \ \ \ 1&  0.18  &  0.&0  & 1.0  &--2.4 ,&\ +0.2   \\  
HEA &\ \  a\ \ \ &\  23&\ \ \ \ o\ \ &\ \ \ \ 2&   0.18 &--0.&066&1.0   &--1.1 ,&\ +0.4   \\  
JON &\ \  d\ \ \ &\  49&\ \ \ \ o\ \ &\ \ \ \ 2&   0.13 &  0.&056&1.0   &--0.5 ,&\ +0.3   \\  
KNK &\ \  c\ \ \ &\  32&\ \ \ \ s\ \ &\ \ \ \ 1&  0.09  &--0.&036& 0.911&--2.1 ,&\ +0.2   \\  
LAI &\ \  b\ \ \ &\  51&\ \ \ \ o\ \ &\ \ \ \ 2&   0.16 &--0.&051& 1.0  &--2.5 ,&\ +0.5   \\  
LBO &\ \  a\ \ \ &\  24&\ \ \ \ o\ \ &\ \ \ \ 2&   0.12 &  0.&093&1.0   &--2.7 ,&\ --0.6  \\             
LCH &\ \  a\ \ \ &\  35&\ \ \ \ s\ \ &\ \ \ \ 2&   0.15 &--0.&058&0.665 &--0.5 ,&\ +0.2   \\
LUB &\ \  a\ \ \ &\  22&\ \ \ \ s\ \ &\ \ \ \ 2&   0.12 &--0.&016&0.945 &--2.8 ,&\ --0.6  \\
MAS &\ \  c\ \ \ &\  39&\ \ \ \ s\ \ &\ \ \ \ 1&  0.12  &--0.&040& 0.630&--1.0 ,&\ +0.2   \\
MCW &\ \  a\ \ \ &\  62&\ \ \ \ o\ \ &\ \ \ \ 1&  0.09  &--0.&062&  1.0 &--0.7 ,&\ +0.2   \\
NHS &\ \  c\ \ \ &\  22&\ \ \ \ s\ \ &\ \ \ \ 2&   0.13 &--0.&089&0.885 &--2.5 ,&\ --1.0  \\
PET &\ \  a\ \ \ &\  26&\ \ \ \ s\ \ &\ \ \ \ 2&   0.12 &  0.&014&1.058 &--3.5 ,&\ --0.5  \\
PSB &\ \  a\ \ \ &\  26&\ \ \ \ n\ \ &\ \ \ \ 1&  0.11  &  0.&0  & 1.0  &--3.2 ,&\ --0.7  \\
PSK &\ \  a\ \ \ &\  29&\ \ \ \ o\ \ &\ \ \ \ 2&   0.14 &--0.&033&1.0   &--3.1 ,&\ --0.9  \\
RMB &\ \  a\ \ \ &\  25&\ \ \ \ o\ \ &\ \ \ \ 2&  0.16  &--0.&064& 1.0  &--2.5 ,&\ --0.7  \\
SIC &\ \ ab\ \ \ &\  19&\ \ \ \ n\ \ &\ \ \ \ 2&   0.15 &  0.&0  &1.0   &--1.8 ,&\ +0.5   \\
THE &\ \  a\ \ \ &\  12&\ \ \ \ n\ \ &\ \ \ \ 1&  0.13  &  0.&0  &  1.0 &--2.9 ,&\ +0.4   \\
TLL &\ \  a\ \ \ &\  22&\ \ \ \ o\ \ &\ \ \ \ 2&   0.09 &--0.&114&1.0   &--2.7 ,&\ --1.3  \\
WOR &\ \  a\ \ \ &\  76&\ \ \ \ o\ \ &\ \ \ \ 2&  0.14  &  0.&033&  1.0 &--2.6 ,&\ +0.5   \\
WAL &\ \  b\ \ \ &\  28&\ \ \ \ s\ \ &\ \ \ \ 2&   0.19 &  0.&055&0.873 &--2.0 ,&\ +0.4   \\
ZAS &\ \  c\ \ \ &\  46&\ \ \ \ s\ \ &\ \ \ \ 2&   0.12 &--0.&063&0.608 &--0.6 ,&\ +0.1   \\
\hline                                          
\end{tabular}                                                                                      
}
\end{table}
\end{scriptsize}

\begin{scriptsize}
\begin{table}                                                              
\centering{
\caption{Codes for calibrated original references.}
\label{calrefer}                                                            
\begin{tabular}{@{}lllll@{}}          
\hline
 Code& Reference  \\
\hline                                
 AAM & Alonso, Arribas \& Mart\'{\i}nez-Roger \shortcite{ALO}        \\ 
 AFG & Axer, Fuhrmann \& Geheren \shortcite{AXE}        \\  
 BAL & Balachandran \shortcite{BAL}        \\ 
 BKP & Beers et al. \shortcite{BEE}        \\ 
 BLL & Blackwell \& Lynas-Gray \shortcite{BLA}        \\ 
 BSL & Brown et al. \shortcite{BRO}        \\  
 CGC & Carretta et al \shortcite{CGR}        \\ 
 CLL & Carney et al. \shortcite{CAR}        \\ 
 EAG & Edvardsson et al. \shortcite{EDW}        \\ 
 GCC & Gratton, Carretta \& Castelli \shortcite{GCC}        \\
 GRJ & Gray \& Johanson \shortcite{GAJ}        \\ 
 GRS & Gratton \& Sneden\shortcite{GRS}        \\
 HEA & Hearnshaw \shortcite{HEA}        \\
 JON & Jones \shortcite{JON}        \\
 KNK & Kunzli et al. \shortcite{KUN}        \\
 LAI & Laird \shortcite{LAI}        \\
 LBO & Luck \& Bond \shortcite{LBO}        \\
 LCH & Luck \& Challener \shortcite{LCH}        \\
 LUB & Luck \& Bond \shortcite{LUB}        \\
 MAS & Marsakov \& Shevelev \shortcite{MAR}        \\
 MCW & McWilliam \shortcite{MCW}        \\
 NHS & Nissen, Hoeg \& Schuster \shortcite{NHS}        \\
 PET & Peterson \shortcite{PET}        \\
 PSB & Pilachowski, Sneden \& Booth \shortcite{PSB}        \\
 PSK & Pilachowski, Sneden \& Kraft \shortcite{PSK}        \\
 RMB & Rebolo, Molaro \& Beckman \shortcite{RMB}        \\
 SIC & Silva \& Cornell \shortcite{SYC}        \\
 TAY & Taylor \shortcite{TAY}        \\
 THE & Th\'evenin \shortcite{THE}        \\
 TLL & Tomkin et al. \shortcite{TLL}        \\
 WAL & Wallerstein \shortcite{WAL}        \\
 WOR & Worthey et al. \shortcite{Worthey1}   \\
 ZAS & Zakhozhaj \& Shaparenko \shortcite{ZAK}        \\
\hline                 
\end{tabular}                                             
}               
\end{table}                                                                
\end{scriptsize}

\section{Atmospheric parameters for field stars}
\label{calculo}

The new set of atmospheric parameters of the stellar library presented
in this paper has been derived in different ways depending on the
original literature sources which were
available. Table~\ref{paramfield} lists the final derived atmospheric
parameters for the field stars of the library. In
Table~\ref{comentrefs} we present a brief explanation of the different
methods and the codes we have used to identify them. A more detailed
description is the following:

\begin{enumerate}

\item If the star is included in the sample of SKC, the three atmospheric
parameters from that paper were kept (coded SKC). This was the case
for a total of 108 stars of our sample.

\item When the star is not included in the sample of SKC but in
$N$ previously calibrated sources, and the original parameters are
within the calibration ranges listed in Tables~\ref{caltemp},
\ref{calgrav} and~\ref{calmetal}, the new parameters $P$ were
determined by taking the weighted average:

\begin{equation}P = \frac{\sum_{i=1}^{N} p^{*}_{i}/ \sigma_{i}^{2}}
{\sum_{i=1}^{N} 1/\sigma_{i}^{2}} 
\end{equation} 
where $p^{\ast}_{i}$ is the corrected parameter and {\large
$\sigma$}$_{i}$ corresponds to the r.m.s. standard deviation of the
comparison with the reference system (SKC, or RF1 \& SKC) (listed in
Tables~\ref{caltemp}, \ref{calgrav} and~\ref{calmetal}). Most (about
60 per cent, see below) of the atmospheric parameters of the stellar
library presented here have been derived in this way (coded RF1 and
RF2).  Note that this procedure was only carried out when the
atmospheric parameter of the star was within the range of calibration,
that is, extrapolations from the fits have never been applied.

\item When the star is not included in any calibrated
source (or, if included, the atmospheric parameters are out of the
calibration range), the final parameter is the raw mean value from
all the available original sources and no previous
correction to the parameter value has been applied. Obviously, these
final parameters should be less reliable than those obtained from
calibrated sources. Since the scatter of effective temperatures from
different sources is different for hot and cold stars than for
intermediate temperature stars, we divided the stars from
non-calibrated sources into three groups:
\begin{description}
\item[--] Stars of intermediate temperatures, 4000~K $<T_{\rm eff}<$ 6300~K 
(coded RF3).
\item[--] Hot stars with $T_{\rm eff} > 6300$~K (coded RF4).
\item[--] Cold stars with $T_{\rm eff} < 4000$~K (coded RF5).
\end{description}
\noindent
This will allow us to derive a more accurate temperature uncertainty
for each one of the three new categories.

\item If there is no data in the literature, both the effective
temperature and surface gravity are estimated from the spectral type
and the luminosity class using the
tabulated atmospheric data from Lang (1991). 
Only a few parameters (2 per cent of the temperature estimations and 7 per cent
for gravities) were derived in this way, which we coded as RF6.

\end{enumerate}

\begin{scriptsize}
\begin{table}                                                              
\centering{
\caption{Brief explanation for the different methods to derive the atmospheric parameters.}
\label{comentrefs}                                                            
\begin{tabular}{@{}ll}          
\hline                                
 SKC & From Soubiran et al. \shortcite{Soubiran} \\ 
 RF1 & From calibrated and corrected sources onto SKC \\  
 RF2 & From calibrated and corrected sources onto RF1 \& SKC\\  
 RF3 & From non calibrated sources. 4000 K $<$ {\it T}$_{\rm eff}$ $<$ 6300 K \\ 
 RF4 & From non calibrated sources. {\it T}$_{\rm eff}$ $>$ 6300 K \\
 RF5 & From non calibrated sources. {\it T}$_{\rm eff}$ $<$ 4000 K \\ 
 RF6 & From spectral type and luminosity class \cite{Lang} \\ 
\hline            
\end{tabular}                                             
}               
\end{table}                                                                
\end{scriptsize}

\begin{figure}
\centerline{
\psfig{figure=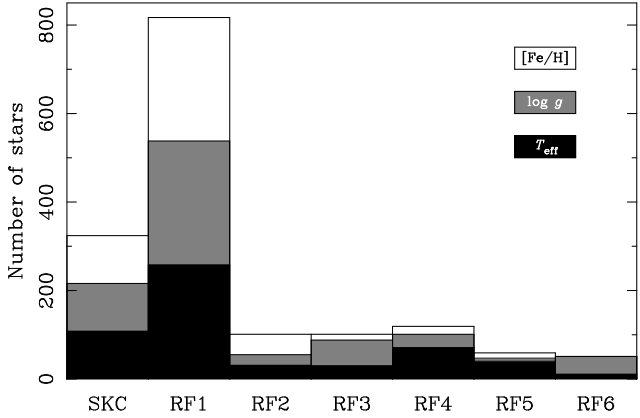}
}  
\caption{Histogram illustrating the total number of stars with
effective temperature (black bars), gravity (grey bars) and
metallicity (white bars) in each category.}
\label{histog}
\end{figure}

To summarize, Fig.~\ref{histog} illustrates the number of stars with
final atmospheric parameters in each different category. A total of
549 temperatures, 547 gravities and 476 metallicities were derived for
the 550 field stars of the stellar library. RF1 is clearly the most
populated category, including about half of the final atmospheric
parameters. Moreover, it is worth noting that most of the effective
temperatures (72.3 per cent), gravities (75.3 per cent) and
metallicities (91.0 per cent) were taken from the initial reference
system (SKC), or derived from calibrated and corrected original
sources (RF1 and RF2).

A detailed table containing all the original data which were used to
derive the final atmospheric parameters of the stellar library is
available from:\\{\tt
http://www.ucm.es/info/Astrof/ellipt/CATRIPLET.html}\\ and\\ {\tt
http://www.nottingham.ac.uk/\~{}ppzrfp/CATRIPLET.html}.

\begin{scriptsize}
\begin{table*}                                                          
\begin{center}
\caption{Final atmospheric parameters of field stars. References for
atmospheric parameters: SKC from Soubiran et al. (1998). Numerical
references $ijk$ indicate that {\it T}$_{{\rm eff}}$ is from RF$i$,
$\log$ g from RF$j$ and [Fe/H] from RF$k$ (see Table~5). Sources for
spectral types are the Bright Star Catalog (Hoffleit \& Jaschek 1982),
Andrillat, Jaschek \& Jaschek (1995), Gorgas et al. (1999), the
Hipparcos Input Catalog and the Simbad database at {\tt
http://simbad.u-strasbg.fr/Simbad}.}
\label{paramfield}
\begin{tabular}{@{}llrlrlccllrlrl}
\hline
HD/Other & SpT & $T_{\rm eff}$ & $\log$ g & [Fe/H]& Ref & & &HD/Other & SpT &
$T_{\rm eff}$ & $\log$ g & [Fe/H]& Ref \\
\hline
   108 & 06 f pec        &  38367 & 3.68&        &  44     & & & 20041 & A0 Ia           &   9480 & 2.13&        &  41     \\
   249 & K1 IV           &   4723 & 2.40& --0.32 &  232    & & & 20630 & G5 V            &   5576 & 4.41&   0.03 &  131    \\
   417 & K0 III          &   4825 & 2.40& --0.32 &  232    & & & 20893 & K3 III          &   4340 & 2.03&   0.08 &  111    \\
  1461 & G0 V            &   5816 & 4.30&   0.20 &  131    & & & 22049 & K2 V            &   5052 & 4.57& --0.15 &{\sc skc}\\
  1918 & G9 III          &   4863 & 2.01& --0.53 &  232    & & & 22484 & F8 V            &   5933 & 4.03& --0.09 &  111    \\
  2665 & G5 IIIwe        &   5013 & 2.35& --1.96 &{\sc skc}& & & 22879 & F9 V            &   5808 & 4.29& --0.83 &  111    \\
  2857 & A2 (HB)         &   7563 & 2.67& --1.60 &  441    & & & 23249 & K0 IV           &   4884 & 3.40& --0.11 &  111    \\
  3443 & K1 V + ...      &   5335 & 4.57& --0.14 &  131    & & &23439 A& K1 V            &   5118 & 4.50& --1.02 &{\sc skc}\\
  3546 & G5 III          &   4942 & 2.73& --0.66 &{\sc skc}& & &23439 B& K2 V            &   4792 & 4.65& --1.02 &  111    \\
  3567 & F5 V            &   5917 & 3.96& --1.32 &{\sc skc}& & & 23841 & K1 III          &   4500 & 1.30& --0.95 &  331    \\
  3651 & K0 V            &   5417 & 4.63&   0.01 &  122    & & & 24451 & K4 V            &   4357 & 4.61&        &  11     \\
  4307 & G0 V            &   5742 & 4.07& --0.25 &  111    & & & 25329 & K1 Vsb          &   4787 & 4.58& --1.72 &{\sc skc}\\
  4614 & G0 V            &   5848 & 4.40& --0.27 &{\sc skc}& & & 26297 & G5-6 IVw        &   4316 & 1.06& --1.67 &{\sc skc}\\
  4628 & K2 V            &   4960 & 4.60& --0.29 &{\sc skc}& & & 26462 & F4 V            &   6814 & 4.12&   0.10 &  412    \\
  4656 & K5 III          &   3915 & 1.45& --0.14 &  111    & & & 26690 & F3 V            &   6925 & 3.96&   0.08 &  411    \\
  5395 & G8 III-IV       &   4797 & 2.55& --0.70 &{\sc skc}& & & 26965 & K1 V            &   5073 & 4.19& --0.31 &{\sc skc}\\
  6186 & G9 III          &   4857 & 2.67& --0.33 &  111    & & & 27295 & B9 IV           &  11677 & 3.93& --0.73 &  444    \\
  6203 & K0 III-IV       &   4492 & 2.60& --0.29 &  111    & & & 27371 & K0 III          &   4961 & 2.71&   0.07 &  111    \\
  6474 & G4 Ia           &   6241 & 1.55&   0.25 &  111    & & & 27697 & K0 III          &   4966 & 2.76&   0.17 &  111    \\
  6695 & A3 V            &   8390 & 4.30&        &  41     & & & 28305 & G9.5 III        &   4844 & 2.68&   0.11 &  111    \\
  6755 & F8 V            &   5102 & 2.40& --1.41 &{\sc skc}& & & 28307 & K0 III          &   4981 & 2.87&   0.10 &  111    \\
  6833 & G8 III          &   4380 & 1.25& --0.99 &{\sc skc}& & & 30455 & G2 V            &   5685 & 4.45& --0.36 &  112    \\
  6860 & M0 III          &   3845 & 1.57&   0.10 &  255    & & & 30649 & G1 V-VI         &   5693 & 4.23& --0.50 &  111    \\
  6903 & G0 III          &   5570 & 2.9 &        &  36     & & & 30652 & F6 V            &   6482 & 4.35&   0.05 &  111    \\
  7010 & K0 IV           &   5000 & 3.3 &        &  66     & & & 30743 & F3-5 V          &   6395 & 4.12& --0.33 &  111    \\
  7927 & F0 Ia           &   7425 & 0.70&        &  44     & & & 34334 & K3 III          &   4211 & 1.96& --0.40 &  111    \\
  8424 & A0 Vnn          &   8455 & 4.10&        &  41     & & & 34411 & G0 V            &   5835 & 4.17&   0.06 &{\sc skc}\\
  9826 & F8 V            &   6135 & 4.08&   0.11 &  111    & & & 35369 & G8 III          &   4863 & 2.50& --0.26 &{\sc skc}\\
 10307 & G2 V            &   5847 & 4.28&   0.02 &  111    & & & 35601 & M1.5 Ia         &   3550 & 0.00& --0.20 &  661    \\
 10380 & K3 III          &   4057 & 1.43& --0.25 &{\sc skc}& & & 35620 & K4 IIIp         &   4367 & 1.75& --0.03 &{\sc skc}\\
 10476 & K1 V            &   5150 & 4.44& --0.17 &{\sc skc}& & & 36003 & K5 V            &   4464 & 4.61&   0.09 &  113    \\
 10700 & G8 V            &   5264 & 4.36& --0.50 &{\sc skc}& & & 36079 & G5 II           &   5170 & 2.04& --0.38 &  111    \\
 10780 & K0 V            &   5393 & 4.60&   0.43 &  122    & & & 36162 & A3 Vn           &   8260 & 4.28&        &  41     \\
 10975 & K0 III          &   4788 & 2.40& --0.30 &  232    & & & 37160 & G8 III-IV       &   4668 & 2.46& --0.50 &{\sc skc}\\
 11004 & F7 V            &   4841 & 2.5 &        &  16     & & & 38393 & F6 V            &   6302 & 4.26& --0.05 &  111    \\
 12014 & K0 Ib           &   5173 & 2.35&   0.45 &  111    & & & 38656 & G8 III          &   4927 & 2.52& --0.22 &  111    \\
 12533 & K3 IIb          &   4383 & 0.92& --0.23 &  333    & & & 38751 & G8 III          &   4748 & 2.27&   0.04 &  111    \\
 12929 & K2 III          &   4458 & 2.24& --0.18 &  111    & & & 39587 & G0 V            &   5869 & 4.45& --0.01 &  111    \\
 13043 & G2 V            &   5695 & 3.68&   0.10 &  111    & & & 39801 & M2 Iab          &   3614 & 0.00&        &  55     \\
 13161 & A5 III          &   8100 & 3.1 &        &  66     & & & 39970 & A0 Ia           &   9400 & 1.43&        &  41     \\
 13267 & B5 Ia           &  13800 & 2.4 &        &  46     & & & 41117 & B2 Iave         &  17482 & 2.70&        &  41     \\
 13611 & G8 Iab          &   5040 & 2.59& --0.23 &  111    & & & 41597 & G8 III          &   4700 & 2.38& --0.54 &{\sc skc}\\
 13783 & G8 V            &   5338 & 4.35& --0.55 &{\sc skc}& & & 41636 & G9 III          &   4708 & 2.50& --0.20 &  111    \\
 13974 & G0 V            &   5700 & 4.42& --0.33 &{\sc skc}& & & 41692 & B5 IV           &  14411 & 3.12& --0.40 &  444    \\
 14134 & B3 Ia           &  15150 & 2.6 &        &  46     & & & 42475 & M1 Iab          &   4000 & 0.70& --0.36 &  665    \\
 14662 & F7 Ib           &   5900 & 1.35& --0.03 &  333    & & & 43318 & F6 V            &   6212 & 3.93& --0.14 &  111    \\
 14802 & G1 V            &   5629 & 3.59& --0.08 &  111    & & & 44007 & G5 IVw          &   4969 & 2.26& --1.47 &{\sc skc}\\
 14938 & F5              &   6132 & 4.03& --0.34 &  111    & & & 45282 & G0              &   5348 & 3.24& --1.44 &{\sc skc}\\
 15596 & G5 III-IV       &   4755 & 2.50& --0.70 &{\sc skc}& & & 46687 & C II            &   2831 &     &   0.20 &  5 5    \\
 16160 & K3 V            &   4718 & 4.50& --0.07 &  131    & & & 46703 & F7 IVw          &   6000 & 0.4 & --1.70 &  333    \\
 16901 & G0 Ib-II        &   5478 & 1.0 &   0.00 &  331    & & & 47205 & K1 IV           &   4753 & 2.93&   0.05 &  111    \\
 17378 & A5 Ia           &   8580 & 1.35&        &  41     & & & 47914 & K5 III          &   3976 & 1.49&   0.05 &  111    \\
 17491 & M4 III          &   3565 & 0.6 &        &  55     & & & 48433 & K1 III          &   4460 & 1.88& --0.25 &{\sc skc}\\
 17548 & F8              &   5944 & 4.28& --0.59 &  111    & & & 48682 & G0 V            &   5946 & 4.07&   0.05 &  111    \\
 17709 & K5 III          &   3894 & 1.14& --0.25 &  111    & & & 49161 & K4 III          &   4180 & 1.46&   0.08 &  111    \\
 18191 & M6 III          &   3289 & 0.3 &        &  55     & & & 49293 & K0 III          &   4629 & 2.16& --0.02 &  111    \\
 18391 & G0 Ia           &   5500 & 0.00& --0.28 &  331    & & & 50778 & K4 III          &   4009 & 1.60& --0.27 &  111    \\
 19373 & G0 V            &   5989 & 4.19&   0.16 &  131    & & & 51440 & K2 III          &   4402 & 2.28& --0.35 &  111    \\
 19445 & A4p             &   5918 & 4.35& --2.05 &{\sc skc}& & & 52005 & K4 Iab          &   4116 & 0.20& --0.20 &  121    \\
 19476 & K0 III          &   4852 & 2.92&   0.10 &{\sc skc}& & & 52973 & G0 Ib var       &   5727 & 1.63&   0.34 &  333    \\
\hline
\end{tabular}
\end{center}
\end{table*}
\end{scriptsize}

\begin{scriptsize}
\begin{table*}                                                          
\begin{center}
\contcaption{}
\begin{tabular}{@{}llrlrlccllrlrl}
\hline
HD/Other & SpT &  $T_{\rm eff}$ & $\log$ g & [Fe/H]&  Ref  & & &HD/Other & SpT &  $T_{\rm eff}$ & $\log$ g & [Fe/H]&  Ref  \\
\hline
 54300 & Spe             &   2700 &     &        &  5      & & & 89025 & F0 III          &   7083 & 3.2 &        &  46     \\
 54716 & K4 Iab          &   4018 & 1.62& --0.16 &  111    & & & 89449 & F6 IV           &   6333 & 4.06&   0.21 &  111    \\
 54719 & K2 III          &   4367 & 1.77&   0.08 &  111    & & & 90508 & G1 V            &   5787 & 4.40& --0.21 &{\sc skc}\\
 54810 & K0 III          &   4697 & 2.35& --0.33 &  112    & & & 93487 & F8              &   5250 & 1.80& --1.05 &  321    \\
 55575 & G0 V            &   5905 & 4.39& --0.28 &  111    & & & 94028 & F4 V            &   5941 & 4.21& --1.49 &{\sc skc}\\
 57060 & 07 Ia           &  35950 & 3.2 &        &  46     & & & 94247 & K3 III          &   4221 & 2.17& --0.16 &  111    \\
 57061 & 09 Ib           &  32300 & 3.0 &        &  46     & & & 94705 & M5.5 III        &   3330 & 0.20&        &  52     \\
 57118 & F0 Ia           &   7700 & 1.7 &        &  66     & & & 95128 & G0 V            &   5834 & 4.34&   0.04 &  111    \\
 57264 & G8 III          &   4620 & 2.72& --0.33 &  111    & & & 95272 & K0 III          &   4637 & 2.33& --0.05 &  111    \\
 58207 & K0 III          &   4786 & 2.55& --0.11 &  111    & & & 97907 & K3 III          &   4350 & 2.07& --0.10 &  111    \\
 58551 & F6 V            &   6145 & 4.18& --0.55 &  111    & & &98230/1& G0 V            &   5831 & 4.46& --0.34 &  333    \\
 58972 & K3 III          &   4031 & 1.81& --0.28 &  111    & & &101501 & G8 V            &   5388 & 4.60& --0.13 &  111    \\
 59612 & A5 Ib           &   8100 & 1.45&   0.08 &  444    & & &102224 & K0 III          &   4383 & 2.02& --0.46 &{\sc skc}\\
 60179 & A1 V            &  10286 & 4.0 &   0.48 &  444    & & &102328 & K3 III          &   4395 & 2.09&   0.35 &  111    \\
 60522 & M0 III-IIIb     &   3854 & 1.20&   0.12 &  111    & & &102634 & F7 V            &   6337 & 4.12&   0.28 &  111    \\
 61603 & K5 III          &   3809 & 1.50&   0.24 &  111    & & &102870 & F8 V            &   6109 & 4.20&   0.17 &{\sc skc}\\
 61913 & M3 II-III       &   3530 & 0.70&        &  66     & & &103095 & G8 Vp           &   5025 & 4.56& --1.36 &{\sc skc}\\
 61935 & K0 III          &   4779 & 2.50& --0.06 &  111    & & &103736 & G8 III          &   4900 & 2.3 &        &  66     \\
 62345 & G8 IIIa         &   5015 & 2.63& --0.08 &  111    & & &103799 & F6 V            &   6174 & 3.85& --0.48 &  111    \\
 62721 & K5 III          &   3960 & 1.51& --0.22 &  111    & & &103877 & Am              &   7306 & 4.0 &   0.40 &  441    \\
 63302 & K3 Iab          &   4500 & 0.2 &   0.12 &  331    & & &104985 & G9 III          &   4667 & 2.20& --0.37 &  232    \\
 63352 & K0 III          &   4226 & 2.20& --0.31 &  111    & & &105546 & G2 IIIw         &   5228 & 2.50& --1.50 &  221    \\
 63700 & G6 Ia           &   4990 & 1.15&   0.24 &  333    & & &106516 & F5 V            &   6153 & 4.36& --0.73 &  111    \\
 64606 & G8 V            &   5210 & 4.24& --0.97 &{\sc skc}& & &107213 & F8 Vs           &   6302 & 4.01&   0.29 &  111    \\
 65583 & G8 V            &   5262 & 4.45& --0.56 &  131    & & &107328 & K0 IIIb         &   4444 & 2.20& --0.33 &  111    \\
 65714 & G8 III          &   4840 & 1.50&   0.27 &  111    & & &107752 & G5              &   4625 & 0.80& --2.74 &  121    \\
 66141 & K2 III          &   4258 & 1.90& --0.30 &  111    & & &107950 & G6III           &   5092 & 2.28& --0.11 &  111    \\
 69267 & K4 III          &   4037 & 1.51& --0.11 &  111    & & &108177 & F5 VI           &   6067 & 4.25& --1.70 &  111    \\
 69897 & F6 V            &   6250 & 4.24& --0.24 &  111    & & &108317 & G0              &   5083 & 2.58& --2.36 &{\sc skc}\\
 70272 & K5 III          &   3897 & 1.28&   0.04 &  111    & & &109995 & A0 V (HB)       &   8034 & 2.98& --1.55 &  444    \\
 72184 & K2 III          &   4627 & 2.61&   0.12 &  111    & & &110184 & G5              &   4380 & 0.63& --2.30 &{\sc skc}\\
 72324 & G9 III          &   4885 & 2.13&   0.16 &  111    & & &110411 & A0 V            &   8970 & 4.36& --1.00 &  444    \\
 72905 & G1.5 Vb         &   5853 & 4.48& --0.08 &  111    & & &110897 & G0 V            &   5830 & 4.23& --0.48 &{\sc skc}\\
 73394 & G5 IIIw         &   4500 & 1.10& --1.38 &  321    & & &111721 & G6 V            &   5014 & 3.22& --1.21 &  111    \\
 73471 & K2 III          &   4488 & 2.00&   0.11 &  111    & & &112014 & A0 V            &   9520 & 4.1 &        &  66     \\
 73593 & G8 IV           &   4717 & 2.25& --0.15 &  112    & & &112028 & A1 III          &   9480 & 3.3 &        &  66     \\
 73665 & K0 III          &   4964 & 2.35&   0.13 &  112    & & &112412 & F0 V            &   6462 & 4.10& --0.11 &  411    \\
 73710 & K0 III          &   4930 & 2.33&   0.28 &  111    & & &112413 & A0 spe          &   9944 & 3.85&   0.32 &  444    \\
 74000 & F6 VI           &   6197 & 4.39& --2.02 &  111    & & &112989 & G9 III          &   4693 & 2.61&   0.20 &  111    \\
 74377 & K3 V            &   4912 & 4.63& --0.07 &  113    & & &113092 & K2 III          &   4280 & 1.94& --0.70 &  111    \\
 74395 & G2 Iab          &   5250 & 1.3 & --0.05 &  331    & & &113139 & F2 V            &   6810 & 3.87&   0.22 &  411    \\
 74442 & K0 III          &   4657 & 2.51& --0.06 &  111    & & &113226 & G8 IIIvar       &   4983 & 2.80&   0.05 &{\sc skc}\\
 74462 & G5 IV           &   4527 & 1.53& --1.40 &  111    & & &113285 & M8 III          &   2924 & 0.00&        &  51     \\
 75732 & G8 V            &   5079 & 4.48&   0.16 &{\sc skc}& & &113848 & F4 V            &   6593 & 3.83& --0.16 &  411    \\
 76932 & F7-8 IV-V       &   5866 & 3.96& --0.93 &{\sc skc}& & &114710 & G0 V            &   5975 & 4.40&   0.09 &{\sc skc}\\
 78418 & G5 IV-V         &   5679 & 4.2 & --0.12 &  131    & & &114762 & F9 V            &   5812 & 4.12& --0.75 &{\sc skc}\\
 79211 & M0 V            &   3769 & 4.71& --0.40 &  515    & & &114946 & G6 V            &   5171 & 3.64&   0.13 &  111    \\
 81797 & K3 II-III       &   4120 & 1.54& --0.06 &  111    & & &114961 & M7 III          &   3014 & 0.00& --0.84 &  512    \\
 82210 & G4 III-IV       &   5208 & 3.19& --0.28 &  111    & & &115043 & G1 V            &   5923 & 4.40& --0.07 &  111    \\
 82328 & F6 IV           &   6311 & 3.90& --0.17 &  111    & & &115444 & K0              &   4736 & 1.70& --2.71 &  121    \\
 82885 & G8 IV--V        &   5487 & 4.61&   0.07 &  122    & & &115604 & F3 III          &   7200 & 3.0 &   0.33 &  441    \\
 83618 & K3 III          &   4231 & 1.74& --0.08 &  111    & & &115617 & G6 V            &   5536 & 4.36& --0.01 &  111    \\
 84441 & G1 II           &   5310 & 1.81& --0.13 &  111    & & &116114 & Ap              &   8040 & 4.17&   0.48 &  441    \\
 84737 & G2V             &   5874 & 4.07&   0.08 &{\sc skc}& & &116842 & A5 V            &   8051 & 4.33&        &  44     \\
 84937 & F5 VI           &   6228 & 4.01& --2.17 &{\sc skc}& & &117176 & G5 V            &   5525 & 3.39& --0.07 &  122    \\
 85503 & K0 III          &   4472 & 2.33&   0.23 &{\sc skc}& & &118055 & K0 IIIw         &   4089 & 0.45& --1.92 &  111    \\
 86728 & G1 V            &   5742 & 4.21&   0.13 &{\sc skc}& & &120136 & F7 V            &   6304 & 4.15&   0.27 &  211    \\
 87140 & K0              &   5099 & 2.76& --1.70 &{\sc skc}& & &120452 & K0.5 III-IIIb   &   4783 & 2.59&   0.03 &  111    \\
 87737 & A0 Ib           &   9959 & 1.98&   0.02 &  444    & & &120933 & K5 III          &   3820 & 1.29&   0.56 &  111    \\
 88230 & K7 V            &   3861 & 4.68& --0.93 &  111    & & &121146 & K2 IV           &   4403 & 3.00& --0.12 &  111    \\
 88284 & K0 III          &   4937 & 2.86&   0.15 &  111    & & &121370 & G0 IV           &   6003 & 3.62&   0.25 &  111    \\
 88609 & G5 IIIwe        &   4513 & 1.26& --2.64 &{\sc skc}& & &121447 & K4 III          &   4200 & 0.8 & --0.05 &  331    \\
 89010 & G2 IV           &   5692 & 3.92&   0.01 &  111    & & &122563 & F8 IV           &   4566 & 1.12& --2.63 &{\sc skc}\\
\hline
\end{tabular}
\end{center}
\end{table*}
\end{scriptsize}

\begin{scriptsize}
\begin{table*}                                                          
\begin{center}
\contcaption{}
\begin{tabular}{@{}llrlrlccllrlrl}
\hline
HD/Other & SpT &  $T_{\rm eff}$ & $\log$ g & [Fe/H]&  Ref  & & &HD/Other & SpT &  $T_{\rm eff}$ & $\log$ g & [Fe/H]&  Ref  \\
\hline
122956 & G6 IV-Vw        &   4635 & 1.49& --1.75 &{\sc skc}& & &147379 B& M3 V           &   3247 & 4.84&        &  51     \\
123299 & A0 III          &  10080 & 3.30& --0.56 &  444    & & &147677 & K0 III          &   4923 & 2.71& --0.01 &  111    \\
123657 & M4 III          &   3452 & 0.6 &   0.00 &  562    & & &148513 & K4 IIIp         &   4014 & 1.67&   0.11 &  111    \\
124186 & K4 III          &   4346 & 2.10&   0.24 &  111    & & &148743 & A7 Ib           &   7100 & 1.60& --0.15 &  441    \\
124547 & K3 III          &   4130 & 1.81&   0.23 &  111    & & &148783 & M6 III          &   3244 & 0.2 &   0.02 &  555    \\
124850 & F7 IV           &   6135 & 3.83& --0.10 &  111    & & &148816 & F9 V            &   5831 & 4.22& --0.73 &  111    \\
124897 & K2 IIIp         &   4361 & 1.93& --0.53 &{\sc skc}& & &149009 & K5 III          &   3853 & 1.60&   0.30 &  211    \\
125454 & G9 III          &   4797 & 2.58& --0.15 &  111    & & &149161 & K4 III          &   3910 & 1.39& --0.17 &  111    \\
125560 & K3 III          &   4381 & 2.06&   0.08 &  111    & & &149414 & G5 V            &   4941 & 4.55& --1.36 &  121    \\
126327 & M7.5 III        &   3000 & 0.00& --0.61 &  512    & & &149661 & K0 V            &   5159 & 4.56&   0.13 &  132    \\
126660 & F7 V            &   6227 & 3.84& --0.27 &  111    & & &150177 & F3 V            &   6019 & 3.99& --0.57 &{\sc skc}\\
126681 & G3 V            &   5565 & 4.78& --1.29 &  111    & & &150275 & K1 III          &   4642 & 2.55& --0.54 &  232    \\
127243 & G3 IV           &   4978 & 3.20& --0.59 &{\sc skc}& & &151203 & M3 IIIab        &   3640 & 0.70&        &  51     \\
127665 & K3 III          &   4259 & 1.83& --0.09 &  111    & & &151217 & K5 III          &   4137 & 1.52& --0.03 &  111    \\
127762 & A7 III          &   7840 & 3.2 &        &  46     & & &152792 & G0 V            &   5612 & 4.12& --0.25 &  111    \\
128167 & F2 V            &   6721 & 4.38& --0.39 &  111    & & &153210 & K2 III          &   4557 & 2.28&   0.05 &  111    \\
129312 & G8 III          &   4880 & 2.45& --0.06 &  111    & & &153597 & F6 Vvar         &   6211 & 4.36& --0.09 &  111    \\
130109 & A0 V            &   9820 & 4.35&        &  44     & & &154783 & Am              &   7782 & 4.05&   0.30 &  411    \\
130694 & K4 III          &   4040 & 1.62& --0.28 &  111    & & &155358 & G0              &   5831 & 4.12& --0.67 &  111    \\
130705 & K4 II-III       &   4335 & 2.10&   0.41 &  111    & & &156014 & M5 Ib-II        &   3293 & 0.76&        &  55     \\
131918 & K4 III          &   3970 & 1.49&   0.28 &  111    & & &157089 & F9 V            &   5785 & 4.12& --0.56 &{\sc skc}\\
131976 & M1 V            &   3506 & 4.73&        &  51     & & &157214 & G0 V            &   5682 & 4.25& --0.39 &{\sc skc}\\
131977 & K4 V            &   4533 & 4.79&   0.02 &  131    & & &157881 & K7 V            &   4065 & 4.50&   0.38 &  231    \\
132142 & K1 V            &   5108 & 4.50& --0.55 &{\sc skc}& & &157910 & G5 III          &   5136 & 1.83& --0.26 &  111    \\
132345 & K3 III-IVp      &   4374 & 1.60&   0.42 &  112    & & &159181 & G2 Iab          &   5250 & 1.60&   0.10 &  331    \\
132475 & F6 V            &   5599 & 3.50& --1.66 &  121    & & &159307 & F8              &   6193 & 3.89& --0.72 &  111    \\
132933 & M0.5 IIb        &   3660 & 0.7 &        &  56     & & &159332 & F6 V            &   6187 & 3.84& --0.19 &  111    \\
134063 & G5 III          &   4881 & 2.34& --0.66 &  232    & & &159561 & A5 III          &   7986 & 3.96&   0.01 &  441    \\
134083 & F5 V            &   6575 & 4.32&   0.00 &  111    & & &160365 & F6 III          &   6070 & 3.0 &        &  46     \\
134169 & G1 Vw           &   5798 & 3.87& --0.91 &{\sc skc}& & &160693 & G0 V            &   5768 & 4.14& --0.61 &{\sc skc}\\
134439 & K0 V            &   4940 & 4.85& --1.49 &  121    & & &161797 & G5 IV           &   5411 & 3.87&   0.16 &{\sc skc}\\
134440 & K3 V-VI         &   4742 & 4.67& --1.47 &  131    & & &161817 & A2 VI (HB)      &   7639 & 2.96& --0.95 &  441    \\
135148 & K0              &   4289 & 0.19& --1.96 &  211    & & &162211 & K2 III          &   4513 & 2.44&   0.05 &  111    \\
135722 & G8 III          &   4847 & 2.56& --0.44 &{\sc skc}& & &162555 & K1 III          &   4650 & 2.49& --0.15 &  111    \\
136028 & K5 III          &   3995 & 1.90&   0.19 &  111    & & &163506 & F2 Ibe          &   6491 & 1.7 & --0.35 &  161    \\
136202 & F8 III-IV       &   6082 & 3.84& --0.08 &  122    & & &163588 & K2 III          &   4434 & 2.33& --0.02 &  111    \\
136479 & K1 III          &   4722 & 2.56&   0.14 &  111    & & &163993 & G8 III          &   5028 & 2.69&   0.03 &  111    \\
136726 & K4 III          &   4156 & 1.91&   0.14 &  111    & & &164058 & K5 III          &   3904 & 1.31& --0.05 &  111    \\
137391 & F0 V            &   7190 & 4.14&   0.28 &  444    & & &164136 & F2 II           &   6693 & 2.70& --0.30 &  444    \\
137471 & M1 III          &   3810 & 1.10&        &  51     & & &164259 & F3 V            &   6737 & 4.00& --0.02 &  411    \\
137759 & K2 III          &   4498 & 2.38&   0.05 &  111    & & &164349 & K0.5 IIb        &   4445 & 1.50&   0.39 &  111    \\
138279 & F5              &   5997 & 2.50& --1.67 &  331    & & &164353 & B5 Ib           &  13493 & 2.4 &        &  46     \\
138481 & K5 III          &   3890 & 1.41&   0.26 &  111    & & &165195 & K3p             &   4471 & 1.11& --2.15 &{\sc skc}\\
139669 & K5 III          &   3917 & 1.41& --0.01 &  111    & & &165401 & G0 V            &   5707 & 4.25& --0.45 &  111    \\
140283 & F3 VI           &   5687 & 3.55& --2.53 &{\sc skc}& & &165760 & G8 III-IV       &   4932 & 2.55& --0.04 &  111    \\
140573 & K2 III          &   4528 & 2.43&   0.17 &  111    & & &165908 & F7 V            &   5928 & 4.24& --0.53 &{\sc skc}\\
141004 & G0 Vvar         &   5915 & 4.10& --0.01 &{\sc skc}& & &166161 & G5              &   4905 & 2.31& --1.25 &{\sc skc}\\
141144 & K0 III          &   4750 & 2.1 &        &  66     & & &166207 & K0 III          &   4764 & 2.20&   0.04 &  232    \\
141680 & G8 III          &   4730 & 2.52& --0.21 &  111    & & &166229 & K2.5 III        &   4529 & 2.32&   0.08 &  111    \\
141714 & G3.5 III        &   5230 & 3.02& --0.28 &  111    & & &166620 & K2 V            &   4944 & 4.47& --0.23 &{\sc skc}\\
142091 & K0 III-IV       &   4796 & 3.22&   0.00 &  111    & & &167006 & M3 III          &   3470 & 0.70&   0.00 &  512    \\
142373 & F9 V            &   5821 & 4.13& --0.41 &{\sc skc}& & &167042 & K1 III          &   4927 & 3.46& --0.19 &  111    \\
142860 & F6 V            &   6249 & 4.16& --0.15 &  111    & & &167768 & G3 III          &   5211 & 1.61& --0.65 &  232    \\
142980 & K1 IV           &   4549 & 2.74&   0.11 &  111    & & &168322 & G8.5 IIIb       &   4805 & 2.17& --0.51 &  232    \\
143107 & K3 III          &   4337 & 1.95& --0.22 &  111    & & &168656 & G8 III          &   5056 & 2.82& --0.14 &  111    \\
143761 & G2 V            &   5762 & 4.23& --0.20 &{\sc skc}& & &168720 & M1 III          &   3810 & 1.10&   0.00 &  515    \\
144585 & G5 V            &   5791 & 3.95&   0.29 &  111    & & &168775 & K2 III          &   4535 & 2.08&   0.03 &  111    \\
144872 & K3 V            &   4739 & 4.65& --0.31 &  112    & & &169191 & K3 III          &   4299 & 2.23& --0.11 &  111    \\
145148 & K0 IV           &   4849 & 3.45&   0.10 &  112    & & &169414 & K2.5 IIIab      &   4415 & 2.47& --0.12 &  111    \\
145328 & K0 III          &   4683 & 3.02& --0.18 &  111    & & &170153 & F7 V            &   6008 & 4.36& --0.33 &  113    \\
145675 & K0 V            &   5264 & 4.66&   0.34 &{\sc skc}& & &170693 & K1.5 III        &   4394 & 2.32& --0.38 &  111    \\
146051 & M0.5 III        &   3847 & 1.4 &   0.32 &  255    & & &171443 & K3 III          &   4191 & 1.83& --0.08 &  111    \\
147379 A& M0 V           &   3720 & 4.67& --1.40 &  511    & & &172167 & A0 V            &   9522 & 3.98& --0.64 &  444    \\
\hline
\end{tabular}
\end{center}
\end{table*}
\end{scriptsize}

\begin{scriptsize}
\begin{table*}                                                          
\begin{center}
\contcaption{}
\begin{tabular}{@{}llrlrlccllrlrl}
\hline
HD/Other & SpT &  $T_{\rm eff}$ & $\log$ g & [Fe/H]&  Ref  & & &HD/Other & SpT &  $T_{\rm eff}$ & $\log$ g & [Fe/H]&  Ref  \\
\hline
172365 & F8 Ib-II        &   5500 & 2.10& --0.64 &  333    & & &195593 & F5 Iab          &   6600 & 1.95&   0.09 &  412    \\
172380 & M4-5 II         &   3421 & 0.55&        &  56     & & &195633 & G0 Vw           &   6000 & 3.78& --0.77 &{\sc skc}\\
172816 & M5.2 III        &   3369 & 0.50&        &  56     & & &195636 & B8              &   5478 & 3.4 & --2.65 &  441    \\
172958 & B8 V            &  11300 & 3.75&        &  41     & & &196758 & K1 III          &   4660 & 2.47& --0.06 &  111    \\
173399 & G5 IV           &   5054 & 2.60& --0.39 &  232    & & &196892 & F6 V            &   5762 & 3.68& --1.12 &  111    \\
173780 & K3 III          &   4400 & 2.34& --0.06 &  111    & & &197076 & G5 V            &   5761 & 4.23&   0.01 &  121    \\
174704 & F1 Vp           &   7412 & 3.5 &   0.60 &  441    & & &198149 & K0 IV           &   5013 & 3.19& --0.19 &{\sc skc}\\
174912 & F8              &   5746 & 4.32& --0.48 &{\sc skc}& & &198183 & B5 V            &  14315 & 4.0 &        &  46     \\
174947 & G8-K0 II        &   4840 & 1.20&   0.39 &  111    & & &198478 & B3 Iae          &  16325 & 2.19& --0.23 &  414    \\
174974 & K1 II           &   4750 & 1.10& --0.15 &  331    & & &199191 & K0 III          &   4759 & 2.53& --0.45 &  232    \\
175305 & G5 III          &   4899 & 2.30& --1.43 &{\sc skc}& & &199478 & B8 Iae          &  10800 & 1.9 &        &  44     \\
175317 & F5-6 IV         &   6594 & 4.12&   0.22 &  211    & & &199580 & K0 III-IV       &   5039 & 3.50& --0.13 &  112    \\
175535 & G7 IIIa         &   5064 & 2.55& --0.09 &  111    & & &199960 & G1 V            &   5773 & 4.19&   0.18 &  111    \\
175545 & K2 III          &   4451 & 2.94&   0.23 &  232    & & &200580 & F9 V            &   5733 & 4.28& --0.67 &  111    \\
175588 & M4 II           &   3483 & 0.6 &        &  56     & & &200779 & K6 V            &   4252 & 4.63&        &  11     \\
175638 & A5 V            &   8150 & 3.90&        &  41     & & &200790 & F8 V            &   5928 & 4.13& --0.12 &{\sc skc}\\
175743 & K1 III          &   4635 & 2.45& --0.12 &  112    & & &201099 & G0              &   5829 & 4.11& --0.51 &  111    \\
175751 & K2 III          &   4680 & 2.49& --0.03 &  111    & & &201381 & G8 III          &   5007 & 2.65& --0.09 &  111    \\
175865 & M5 III          &   3420 & 0.50&   0.14 &  515    & & &201626 & G9p             &   4941 & 2.00& --1.50 &  331    \\
176301 & B7 III-IV       &  13100 & 3.50&        &  41     & & &201891 & F8 V-VI         &   5854 & 4.45& --1.11 &{\sc skc}\\
176411 & K2 III          &   4718 & 2.51&   0.06 &  111    & & &203344 & K0 III-IV       &   4658 & 2.33& --0.17 &  111    \\
178717 & K3.5 III        &   4308 & 1.0 & --0.30 &  331    & & &204587 & M0 V            &   4034 & 4.67&        &  11     \\
180711 & G9 III          &   4800 & 2.67& --0.12 &{\sc skc}& & &204771 & K0 III          &   4917 & 2.60& --0.05 &  111    \\
180928 & K4 III          &   4000 & 1.30& --0.38 &  512    & & &205153 & G0 IV           &   5961 & 3.20& --0.01 &  222    \\
181615 & B2 Vpe + ...    &   6545 & 3.9 &   0.48 &  464    & & &205435 & G8 III          &   4989 & 2.76& --0.26 &  111    \\
181984 & K3 III          &   4447 & 2.20&   0.21 &  111    & & &205512 & K1 III          &   4634 & 2.57&   0.03 &{\sc skc}\\
182293 & K3 IVp          &   4505 & 3.00&   0.19 &  232    & & &205650 & F6 V            &   5665 & 3.48& --1.26 &  111    \\
182572 & G8 IVvar        &   5570 & 4.19&   0.31 &{\sc skc}& & &206078 & G8 III          &   4667 & 2.87& --0.46 &  232    \\
182762 & K0 III          &   4820 & 2.78& --0.14 &  111    & & &206165 & B2 Ib           &  17760 & 2.66& --0.33 &  444    \\
182835 & F2 Ib           &   7350 & 2.15&   0.09 &  444    & & &207076 & M7 III          &   3008 & 0.00&        &  51     \\
184406 & K3 III          &   4520 & 2.41&   0.01 &{\sc skc}& & &207134 & K3 III          &   4403 & 2.74&   0.13 &  232    \\
184492 & G8 IIIa         &   4529 & 2.11& --0.04 &  111    & & &207260 & A2 Ia           &   9100 & 2.09&        &  41     \\
184499 & G0 V            &   5738 & 4.02& --0.66 &{\sc skc}& & &207673 & A2 Ib           &   9071 & 1.40&   0.16 &  444    \\
185144 & K0 V            &   5260 & 4.55& --0.24 &{\sc skc}& & &207978 & F6 IV-Vwv       &   6244 & 4.00& --0.62 &  111    \\
185018 & K0 V            &   5550 & 1.3 &        &  66     & & &208501 & B8 Ib           &  12200 & 2.2 &        &  46     \\
185644 & K1 III          &   4536 & 2.67&   0.08 &  232    & & &208906 & F8 V-VI         &   5965 & 4.20& --0.74 &{\sc skc}\\
185859 & B0.5 Iae        &  22780 & 2.80&        &  41     & & &208947 & B2 V            &  20559 & 3.9 &        &  46     \\
186408 & G2 V            &   5815 & 4.30&   0.09 &{\sc skc}& & &209481 & 09 V            &  36300 & 3.9 &        &  46     \\
186427 & G5 V            &   5762 & 4.43&   0.07 &{\sc skc}& & &209975 & 08 Ib           &  29647 & 3.30&   0.30 &  444    \\
186486 & G8 III          &   4994 & 2.88& --0.06 &  111    & & &210027 & F5 V            &   6413 & 4.16&   0.00 &  111    \\
186568 & B8 III          &  10609 & 3.4 &        &  46     & & &210295 & G8              &   4746 & 1.50& --1.42 &  131    \\
186791 & K3 II           &   4187 & 1.40& --0.23 &  111    & & &210745 & K1.5 Ib         &   4500 & 0.75&   0.22 &  333    \\
187299 & G5 Ia           &   5010 & 1.10&   0.15 &  161    & & &210855 & F8 V            &   6199 & 3.78&   0.12 &  111    \\
187691 & F8 V            &   6107 & 4.30&   0.11 &{\sc skc}& & &210939 & K1 III          &   4443 & 2.30&   0.04 &  111    \\
187923 & G0 V            &   5662 & 4.21& --0.09 &  112    & & &211391 & G8 III          &   4943 & 2.70&   0.08 &  111    \\
188056 & K3 III          &   4244 & 2.01&   0.17 &  111    & & &212496 & G8.5 IIIb       &   4696 & 2.72& --0.34 &  111    \\
188310 & G9 IIIb         &   4635 & 2.51& --0.24 &  111    & & &212943 & K0 III          &   4586 & 2.81& --0.34 &{\sc skc}\\
188510 & G5 Vwe          &   5490 & 4.69& --1.59 &{\sc skc}& & &213470 & A3 Ia           &   8800 & 1.38&        &  41     \\
188512 & G8 IVvar        &   5041 & 3.04& --0.04 &{\sc skc}& & &214376 & K2 III          &   4580 & 2.48&   0.20 &  111    \\
188727 & G5 Ib var       &   5684 & 1.60&   0.00 &  113    & & &215257 & F8              &   5871 & 4.30& --0.71 &  111    \\
190360 & G6 IV + M6 V    &   5594 & 3.89&   0.25 &  111    & & &215373 & K0 III          &   4905 & 2.65&   0.03 &  111    \\
190406 & G1 V            &   5821 & 4.10& --0.03 &  232    & & &215648 & F7 V            &   6169 & 4.02& --0.30 &  111    \\
190603 & B1.5 Iae        &  19250 & 2.41&        &  41     & & &216131 & G8 III          &   5018 & 2.78& --0.09 &{\sc skc}\\
190608 & K2 III          &   4795 & 2.63&   0.01 &  111    & & &216143 & G5              &   4496 & 1.27& --2.15 &{\sc skc}\\
191046 & K0 III          &   4345 & 2.01& --0.63 &  232    & & &216174 & K1 III          &   4390 & 2.23& --0.53 &{\sc skc}\\
192310 & K0 Vvar         &   5045 & 4.50&   0.08 &  131    & & &216228 & K0 III          &   4768 & 2.49&   0.01 &{\sc skc}\\
192422 & B0.5 Ib         &  22600 & 2.8 &        &  46     & & &216385 & F7 IV           &   6179 & 3.98& --0.35 &{\sc skc}\\
192947 & G6-8 III        &   5000 & 2.82& --0.08 &  111    & & &217476 & G4 Ia           &   5100 & 0.00&   0.00 &  111    \\
193370 & F6 Ib           &   6200 & 1.59&   0.02 &  333    & & &217877 & F8 V            &   5866 & 4.02& --0.19 &  111    \\
193901 & F7 V            &   5713 & 4.39& --1.11 &  111    & & &218029 & K3 III          &   4290 & 2.05&   0.13 &  111    \\
194598 & F7 V-VI         &   5887 & 4.27& --1.22 &{\sc skc}& & &218031 & K0 IIIb         &   4647 & 2.52& --0.14 &  111    \\
194839 & B0.5 Ia         &  23500 & 2.8 &        &  46     & & &218329 & M1 IIIab        &   3810 & 1.10&        &  51     \\
195592 & O9.5 Ia         &  29000 & 2.8 &        &  46     & & &218470 & F5 V            &   6529 & 4.11& --0.10 &  211    \\
\hline
\end{tabular}
\end{center}
\end{table*}
\end{scriptsize}

\begin{scriptsize}
\begin{table*}                                                          
\begin{centering}
\contcaption{}
\begin{tabular}{@{}llrlrlccllrlrl}
\hline
HD/Other & SpT &  $T_{\rm eff}$ & $\log$ g & [Fe/H]&  Ref  & & &HD/Other & SpT &  $T_{\rm eff}$ & $\log$ g & [Fe/H]&  Ref  \\
\hline
218502 & F3w             &   6030 & 3.76& --1.84 &{\sc skc}& & &BD+ 09 3223    & III             &   5274 & 2.00& --2.23 &  222    \\
218658 & G2 III          &   5160 & 2.62&   0.02 &  111    & & &BD+ 17 4708    & F8 VI           &   6005 & 4.01& --1.74 &{\sc skc}\\
218857 & G6w             &   5082 & 2.41& --1.93 &{\sc skc}& & &BD+ 18 5215    & F5              &   6290 & 4.49& --0.40 &  331    \\
219134 & K3 Vvar         &   4717 & 4.50&   0.05 &{\sc skc}& & &BD+ 19 5116 A  & M4 V            &   3200 & 4.91&        &  51     \\
219449 & K0 III          &   4578 & 2.39& --0.09 &{\sc skc}& & &BD+ 19 5116 B  & M6 V            &   2950 & 5.06&        &  51     \\
219615 & G9 III          &   4830 & 2.57& --0.42 &{\sc skc}& & &BD+ 26 3578    & B5              &   6165 & 4.06& --2.25 &{\sc skc}\\
219617 & F8w             &   5878 & 4.04& --1.39 &{\sc skc}& & &BD+ 30 2034    & K3 III          &   4500 & 0.40& --1.40 &  321    \\
219623 & F7 V            &   6155 & 4.17& --0.04 &{\sc skc}& & &BD+ 30 2611    & G8 III          &   4311 & 0.94& --1.36 &{\sc skc}\\
219734 & M2 III          &   3730 & 0.90&   0.27 &  515    & & &BD+ 34 2476    & A4              &   6231 & 3.80& --2.10 &  111    \\
219945 & K0 III          &   4762 & 2.61& --0.16 &  111    & & &BD+ 41 3306    & K0 V            &   4913 & 4.5 & --0.79 &  131    \\
219962 & K1 III          &   4605 & 2.14& --0.11 &  232    & & &BD+ 43 0044 B  & M6 V            &   3721 & 5.08& --1.40 &  511    \\
219978 & K4.5 Ib         &   4250 & 0.80& --0.15 &  331    & & &BD+ 44 2051 A  & M2 V            &   3544 & 4.85& --1.40 &  511    \\
220009 & K2 III          &   4416 & 2.24& --0.56 &  111    & & &BD+ 52 1601    & G5 IIIw         &   4893 & 2.00& --1.30 &  331    \\
220321 & K0 III          &   4502 & 2.39& --0.35 &  111    & & &BD+ 56 1458    & K7 V            &   4069 & 4.70& --0.18 &  115    \\
221148 & K3 IIIvar       &   4643 & 3.05&   0.40 &  112    & & &BD+ 58 1218    & F8              &   4957 & 1.10& --2.59 &  121    \\
221830 & F9 V            &   5688 & 4.16& --0.44 &{\sc skc}& & &BD+ 59 2723    & F2              &   6111 & 4.25& --2.03 &  131    \\
222107 & G8 III          &   4600 & 2.88& --0.54 &  111    & & &BD+ 61  154    & Be              &        &     &        &         \\
222368 & F7 V            &   6136 & 4.12& --0.14 &  111    & & &BD+ 61 2575    & F8 Ib           &   6241 & 1.85&   0.35 &  111    \\
224930 & G3 V            &   5305 & 4.49& --0.75 &{\sc skc}& & &BD-- 01 2582   & F0              &   5067 & 2.12& --2.32 &  121    \\
231195 & F5 Ia           &   7500 & 1.4 &        &  46     & & &G 275--4       & G               &   6010 & 4.05& --3.45 &  111    \\
232078 & K4-5 III-II     &   3996 & 0.30& --1.73 &  521    & & &G 64--12       & F0              &   6312 & 4.21& --3.35 &  111    \\
232979 & K8 V            &   3769 & 4.70& --0.33 &  515    & & &Gl 699         & M5 V            &   3201 & 5.00& --0.90 &  515    \\
BD+ 01 2916    & K0      &   4442 & 1.10& --1.50 &  321    & & &Gl 725A        & M4 V            &   3451 & 4.8 &        &  56     \\
BD+ 02 3375    & A5      &   5978 & 4.04& --2.35 &{\sc skc}& & &Gl 725B        & M5 V            &   3304 & 4.9 &        &  56     \\
BD+ 04 2621    & G0      &   4607 & 1.10& --2.55 &  321    & & &Luyton 789--6  & M7e             &   2747 & 5.09&        &  51     \\
BD+ 09 3063    &         &   4628 & 1.55& --0.75 &  311    & & &Ross 248       & M6e             &   2799 & 5.12&        &  51     \\
\hline
\end{tabular}
\medskip
\end{centering}

\end{table*}
\end{scriptsize}

\section{Uncertainty estimates in the derived parameters}
\label{errerrs}

In this section, we calculate mean error estimates for the final
atmospheric parameters of the library stars.

In general, the number of original atmospheric parameters for a single
star is not large enough to estimate a reliable individual error. This
is the reason why we present an uncertainty estimate for each one of
the categories defined in Section~\ref{calculo} by calculating a mean
r.m.s standard deviation of all those stars whose final parameters
were derived from two or more original sources. Since categories RF1
and RF2 were derived from corrected atmospheric parameters in an
homogeneous way, it is expected that their uncertainties are very
similar. Therefore, we estimate a more reliable error for both samples
together by using all the stars in RF1 and RF2. If $P_{k}$ is the
final parameter of the $k^{\rm th}$ star derived from $N_{k}$
corrected parameters $p_{i,k}^{*}$, and {\large $\sigma$}$_{i,k}$ is
the r.m.s standard deviation of the fit used to calculate each
corrected parameter, then the unbiased unbiased r.m.s standard
deviation {\large $\sigma$} of both categories is derived as follows:

\begin{equation}
\sigma = \sqrt{\frac{(\sum_{k=1}^{N_{\rm star}} N_{k})(\sum_{k=1}^{N_{\rm star}}
\sum_{i=1}^{N_{k}} (p^{*}_{i,k} - P_{k})^{2}/ \sigma_{i,k}^{2})}
{[\sum_{k=1}^{N_{\rm star}}(N_{k} - 1)](\sum_{k=1}^{N_{\rm star}}
\sum_{i=1}^{N_{k}}1/\sigma_{i,k}^{2})} }
\end{equation}
where $N_{\rm star}$ is the total number of stars with at least two
original sources which were used to derive the final atmospheric
parameter.

Categories RF3, RF4 and RF5 consist of stars in different temperature
ranges, and we have estimated errors for each one of them
separately. The method is the same for the three samples. Using the
same notation and requirements given above and considering
non-corrected original parameters $p_{i,k}$, the mean r.m.s. standard
deviation of each category can be given by:

\begin{equation}
\sigma = \sqrt{\frac{\sum_{k=1}^{N_{\rm star}}\sum_{i=1}^{N_{k}} (p_{i,k} -
P_{k})^{2}}{\sum_{k=1}^{N_{\rm star}}(N_{k} - 1)}}
\end{equation}

It is clear that errors for those categories which were defined from a
single source, namely SKC and RF6, could not be calculated in this
way. However, an error estimation for SKC data is published in the
original paper. They adopt 100 K, 0.5 dex, and 0.3 dex respectively
for {\it T}$_{\rm eff}$, log\,{\it g} and [Fe/H] as an upper limit for
the scatter inherent to their method.

In summary, Table~\ref{errores} shows the uncertainty estimates
obtained for the rest of categories.

\begin{small}
\begin{table}                                                              
\centering{                                                                
\caption{Estimated uncertainties for the new categories.
Columns are: Categories, number of stars with at least two
original references, total number of original references used and mean
r.m.s standard deviation error for each one of the three atmospheric
parameters.}
\label{errores}                                                            
\begin{tabular}{@{}crrr@{}l@{}}          
\hline  
 Category & $N_{\rm star}$ &  $N_{\rm ref}$ &\multicolumn{2}{c}{{\large $\sigma$}$_{T_{\rm eff}}$} \\                   
          &           &            &\multicolumn{2}{c}{{\large $\sigma$}$_{\log g}$}      \\                   
          &           &            &\multicolumn{2}{c}{{\large $\sigma$}$_{{\rm [Fe/H]}}$}\\ 
\hline
RF1 \& RF2&     179   &  492       &   60.&9          \\                   
          &     153   &  413       &    0.&18         \\                   
          &     171   &  519       &    0.&09         \\
RF3       &       7   &   16       &  117.&6          \\                   
          &      13   &   33       &    0.&21         \\                   
          &       3   &    7       &    0.&10         \\
RF4       &      27   &   93       &  721.&4          \\                   
          &      13   &   49       &    0.&32         \\                   
          &       8   &   39       &    0.&29         \\
RF5       &      11   &   34       &  112.&8          \\                   
          &       1   &    4       &    0.&21         \\                   
          &       0   &    0       &    $-$&          \\ 
\hline            
\end{tabular}                                                              
}                                                                          
\end{table}                                                                
\end{small}

\section{Atmospheric parameters for cluster stars}

The stellar library presented in this series of papers is an extension
of the Lick/IDS (Image Dissector Scanner) stellar sample (G93,
W94). This library included a large number of (open and globular)
cluster stars which have been retained in the present version. In this
section we revise the atmospheric parameters of these cluster stars.
The final adopted parameters are presented in Table~\ref{paramcum}.

\begin{scriptsize}
\begin{table*}                                                          
\begin{center}
\caption{Final atmospheric parameters of cluster stars. References for
effective temperatures: (1) Derived from $B-V$ versus $T_{\rm eff}$ relations in ALO (Alonso
et al. 1996b, Alonso et al. 1999); (2) Mean from $B-V$ and $V-K$ versus $T_{\rm eff}$
relations in ALO; (3) From Worthey et al. \shortcite{Worthey1}. Surface
gravities from Gorgas et al. \shortcite{Gorgas2}, Worthey et
al. \shortcite{Worthey1} and references therein. See Table~\ref{tab_zc} for
metallicity sources.
Sources for spectral types of Coma and Hyades
stars are as in Table~\ref{paramfield}. For the rest of clusters, we
list positions in the HR diagram (SGB: subgiant branch; GB: giant
branch; HB: horizontal branch; AGB: asymptotic giant branch).}
\label{paramcum}
\begin{tabular}{@{}llrlrcccllrlrc}
\hline
Name      & SpT &  $T_{\rm eff}$ & $\log$ g & [Fe/H]&  Ref & & &Name     & SpT &  $T_{\rm eff}$ & $\log$ g & [Fe/H]&  Ref  \\
\hline
Coma A   3     &   G9 V       &  4974 &4.530 &--0.05 &1& & &      M67 F 117      &   SGB        &  5353 &3.790 &--0.09 &2\\
Coma A  13     &   K0 V       &  5284 &4.540 &--0.05 &1& & &      M67 F 119      &   SGB        &  6095 &3.940 &--0.09 &1\\
Coma A  14     &   G4 V       &  5224 &4.320 &--0.05 &1& & &      M67 F 125      &   SGB        &  6134 &4.340 &--0.09 &1\\
Coma A  21     &   G7 V       &  5110 &4.410 &--0.05 &1& & &      M67 F 164      &    HB        &  4699 &2.220 &--0.09 &2\\
Coma T  65     &   G0 V       &  5918 &4.300 &--0.05 &1& & &      M67 F 170      &    GB        &  4289 &1.830 &--0.09 &2\\ 
Coma T  68     &   A6 IV-V    &  7905 &4.090 &--0.05 &1& & &      M67 F 175      &   SGB        &  6055 &4.340 &--0.09 &1\\ 
Coma T  82     &   A9 V       &  7352 &4.130 &--0.05 &1& & &      M67 F 193      &    GB        &  4928 &3.300 &--0.09 &2\\ 
Coma T  85     &   G1 V       &  5918 &4.380 &--0.05 &1& & &      M67 F 224      &    GB?       &  4704 &2.530 &--0.09 &2\\ 
Coma T  86     &   F6 V       &  6402 &4.270 &--0.05 &1& & &      M67 F 231      &    GB        &  4850 &2.950 &--0.09 &2\\ 
Coma T  90     &   F5 V       &  6359 &4.280 &--0.05 &1& & &      M67 I-17       &    GB        &  4952 &3.370 &--0.09 &2\\ 
Coma T  97     &   F9 V       &  6032 &4.340 &--0.05 &1& & &      M67 II-22      &   SGB        &  5042 &3.650 &--0.09 &2\\ 
Coma T 102     &   G1 V       &  5844 &4.360 &--0.05 &1& & &      M67 IV-20      &    GB        &  4722 &2.750 &--0.09 &2\\ 
Coma T 114     &   F8 V       &  6446 &4.300 &--0.05 &1& & &      M67 IV-68      &   SGB        &  5158 &3.730 &--0.09 &2\\ 
Coma T 132     &   G5 V       &  5567 &4.470 &--0.05 &1& & &      M67 IV-77      &   SGB        &  4946 &3.570 &--0.09 &2\\ 
Coma T 150     &   G9 V       &  5254 &4.300 &--0.05 &1& & &      M67 IV-81      &   SGB        &  5352 &3.760 &--0.09 &2\\ 
Hya vB  10     &   G0 V       &  5954 &4.410 &  0.13 &1& & &      M71 1-09       &   AGB        &  4672 &1.670 &--0.70 &1\\ 
Hya vB  15     &   G3 V       &  5640 &4.340 &  0.13 &2& & &      M71 1-21       &    GB        &  4364 &1.460 &--0.70 &2\\ 
Hya vB  17     &   G5 V       &  5544 &4.530 &  0.13 &2& & &      M71 1-34       &    HB        &  5075 &2.470 &--0.70 &1\\ 
Hya vB  19     &   F8 V       &  6271 &4.270 &  0.13 &1& & &      M71 1-37       &    GB        &  4574 &2.180 &--0.70 &1\\ 
Hya vB  21     &   K0 V       &  5227 &4.570 &  0.13 &1& & &      M71 1-39       &    HB        &  4976 &2.430 &--0.70 &1\\ 
Hya vB  26     &   G9 V       &  5439 &4.500 &  0.13 &1& & &      M71 1-41       &    HB        &  5123 &2.480 &--0.70 &1\\ 
Hya vB  31     &   G0 V       &  6030 &4.310 &  0.13 &1& & &      M71 1-53       &    GB        &  4167 &1.420 &--0.70 &1\\ 
Hya vB  35     &   F5 V       &  6576 &4.250 &  0.13 &1& & &      M71 1-59       &    GB        &  4623 &2.440 &--0.70 &1\\ 
Hya vB  36     &   F6 V       &  6576 &4.240 &  0.13 &1& & &      M71 1-63       &   AGB        &  4689 &1.820 &--0.70 &1\\ 
Hya vB  37     &   F5 V       &  6708 &4.180 &  0.13 &2& & &      M71 1-64       &    GB        &  4275 &1.510 &--0.70 &1\\ 
Hya vB  63     &   G1 V       &  5772 &4.220 &  0.13 &1& & &      M71 1-65       &    GB        &  4606 &2.200 &--0.70 &1\\ 
Hya vB  64     &   G2 V       &  5689 &4.390 &  0.13 &2& & &      M71 1-66       &   AGB        &  4465 &1.470 &--0.70 &1\\ 
Hya vB  73     &   G2 V       &  5886 &4.350 &  0.13 &2& & &      M71 1-71       &    GB        &  4404 &1.810 &--0.70 &1\\ 
Hya vB  81     &   F6 V       &  6441 &4.300 &  0.13 &1& & &      M71 1-73       &    GB        &  4793 &2.520 &--0.70 &1\\ 
Hya vB  87     &   G8 V       &  5439 &4.480 &  0.13 &1& & &      M71 1-75       &              &  4790 &2.560 &--0.70 &2\\ 
Hya vB  95     &   A8 V n     &  7578 &3.790 &  0.13 &2& & &      M71 1-87       &              &  5075 &2.470 &--0.70 &1\\ 
Hya vB 103     &   F0 V       &  7228 &4.050 &  0.13 &1& & &      M71 1-95       &   AGB        &  4639 &1.670 &--0.70 &1\\ 
Hya vB 104     &   A6 V n     &  8380 &3.870 &  0.13 &3& & &      M71 1-107      &   AGB        &  4919 &1.930 &--0.70 &1\\ 
Hya vB 111     &   F0 V       &  7573 &4.030 &  0.13 &1& & &      M71 1-109      &    GB        &  4723 &2.570 &--0.70 &1\\ 
Hya vB 112     &   Am         &  7888 &4.150 &  0.13 &1& & &      M71 A2         &    HB        &  4840 &2.370 &--0.70 &2\\ 
Hya vB 126     &   F3 IV      &  7339 &4.250 &  0.13 &1& & &      M71 A4         &   AGB        &  4040 &0.740 &--0.70 &2\\ 
Hya vB 140     &   G5 V       &  5377 &4.480 &  0.13 &1& & &      M71 A9         &    GB        &  4151 &1.390 &--0.70 &2\\ 
M10 II-76      &   AGB        &  4623 &1.470 &--1.41 &1& & &      M71 C          &    HB        &  4892 &2.390 &--0.70 &2\\ 
M10 III-85     &   GB         &  4397 &1.200 &--1.41 &1& & &      M71 S          &    GB        &  4247 &1.390 &--0.70 &2\\ 
M13 A 171      &   AGB        &  4566 &1.070 &--1.39 &1& & &      M71 X          &    HB        &  5170 &2.490 &--0.70 &2\\ 
M13 B 786      &   GB         &  3891 &0.620 &--1.39 &1& & &      M71 KC 147     &              &  4901 &2.640 &--0.70 &1\\ 
M13 B 818      &   AGB        &  5301 &1.890 &--1.39 &1& & &      M71 KC 169     &              &  5014 &2.440 &--0.70 &1\\ 
M3  398        &    GB        &  4541 &1.440 &--1.34 &1& & &      M71 KC 263     &              &  4883 &2.660 &--0.70 &1\\ 
M3  III-28     &    GB        &  4093 &0.730 &--1.34 &2& & &      M92 I-10       &    HB        &  9290 &3.440 &--2.16 &3\\ 
M3  IV-25      &    GB        &  4367 &1.210 &--1.34 &2& & &      M92 I-13       &    HB        &  5641 &2.220 &--2.16 &1\\ 
M5  I-45       &    HB        &  5758 &2.610 &--1.11 &1& & &      M92 II-23      &    HB        &  7510 &3.050 &--2.16 &3\\ 
M5  II-51      &    GB        &  4627 &1.690 &--1.11 &2& & &      M92 III-13     &    GB        &  4178 &0.580 &--2.16 &2\\ 
M5  II-53      &    HB        & 10460 &3.660 &--1.11 &3& & &      M92 IV-114     &    GB        &  4728 &1.580 &--2.16 &2\\ 
M5  II-76      &    HB        &  5974 &2.700 &--1.11 &1& & &      M92 VI-74      &    HB        &  5752 &2.210 &--2.16 &1\\ 
M5  III-03     &    GB        &  4031 &0.660 &--1.11 &2& & &      M92 IX-12      &   AGB        &  5677 &1.930 &--2.16 &1\\ 
M5  IV-19      &    GB        &  4113 &0.840 &--1.11 &2& & &      M92 XII-8      &    GB        &  4477 &1.000 &--2.16 &2\\ 
M5  IV-59      &    GB        &  4245 &0.850 &--1.11 &2& & &      M92 XII-24     &    HB        & 11100 &3.750 &--2.16 &3\\ 
M5  IV-86      &    HB        &  5576 &2.460 &--1.11 &2& & &      NGC 188 I-55   &   SGB        &  5375 &3.910 &--0.05 &1\\ 
M5  IV-87      &    HB        &  5864 &2.630 &--1.11 &1& & &      NGC 188 I-57   &    GB        &  4740 &3.070 &--0.05 &1\\ 
M67 F 084      &    HB        &  4733 &2.320 &--0.09 &2& & &      NGC 188 I-61   &    GB        &  4915 &3.350 &--0.05 &1\\ 
M67 F 094      &   SGB        &  6219 &4.070 &--0.09 &2& & &      NGC 188 I-69   &    GB        &  4427 &2.350 &--0.05 &1\\ 
M67 F 105      &    GB        &  4461 &2.230 &--0.09 &2& & &      NGC 188 I-75   &    GB        &  4895 &3.220 &--0.05 &1\\ 
M67 F 108      &    GB        &  4255 &1.830 &--0.09 &2& & &      NGC 188 I-85   &    GB        &  4820 &3.550 &--0.05 &1\\ 
M67 F 115      &   SGB        &  6004 &3.890 &--0.09 &2& & &      NGC 188 I-88   &   SGB        &  5195 &3.850 &--0.05 &1\\ 
\hline
\end{tabular}
\end{center}
\end{table*}
\end{scriptsize}

\begin{scriptsize}
\begin{table*}                                                          
\begin{center}
\contcaption{}
\begin{tabular}{@{}llrlrcccllrlrc}
\hline
Name     & SpT &  $T_{\rm eff}$ & $\log$ g & [Fe/H]&  Ref  & & &Name     & SpT &  $T_{\rm eff}$ & $\log$ g & [Fe/H]&  Ref  \\
\hline
NGC 188 I-97   &   SGB        &  5110 &3.820 &--0.05 &1& & &    NGC 188 II-181 &    GB        &  4300 &2.190 &--0.05 &1\\
NGC 188 I-105  &    HB        &  4613 &2.190 &--0.05 &1& & &    NGC 188 II-187 &    GB        &  4936 &3.330 &--0.05 &1\\
NGC 188 I-116  &              &  5148 &3.230 &--0.05 &1& & &    NGC 6171 04    &    HB        &  6039 &2.750 &--0.95 &1\\
NGC 188 II-52  &   SGB        &  5501 &3.940 &--0.05 &1& & &    NGC 6171 45    &    HB        &  5856 &2.840 &--0.95 &1\\
NGC 188 II-64  &   SGB        &  5808 &4.090 &--0.05 &1& & &    NGC 7789 415   &    GB        &  3885 &1.060 &--0.24 &1\\ 
NGC 188 II-67  &   SGB        &  5956 &4.130 &--0.05 &1& & &    NGC 7789 468   &    GB        &  4228 &1.600 &--0.24 &1\\
NGC 188 II-69  &   SGB        &  6032 &4.170 &--0.05 &1& & &    NGC 7789 501   &    GB        &  4102 &1.380 &--0.24 &1\\
NGC 188 II-72  &    GB        &  4410 &2.410 &--0.05 &1& & &    NGC 7789 575   &    GB        &  4506 &1.980 &--0.24 &1\\
NGC 188 II-76  &    HB        &  4578 &2.180 &--0.05 &1& & &    NGC 7789 669   &    GB        &  4214 &1.570 &--0.24 &1\\ 
NGC 188 II-79  &   SGB        &  5055 &3.550 &--0.05 &1& & &    NGC 7789 676   &    HB        &  4961 &2.320 &--0.24 &1\\ 
NGC 188 II-88  &    GB        &  4543 &2.710 &--0.05 &1& & &    NGC 7789 859   &    GB        &  4625 &2.270 &--0.24 &1\\ 
NGC 188 II-93  &   SGB        &  5469 &3.930 &--0.05 &1& & &    NGC 7789 875   &    HB        &  4921 &2.360 &--0.24 &1\\ 
NGC 188 II-122 &    GB        &  4936 &3.410 &--0.05 &1& & &    NGC 7789 897   &    HB        &  4921 &2.350 &--0.24 &1\\ 
NGC 188 II-126 &    GB        &  4936 &3.450 &--0.05 &1& & &    NGC 7789 971   &    GB        &  3860 &1.030 &--0.24 &1\\ 
\hline
\end{tabular}
\end{center}
\end{table*}
\end{scriptsize}

\subsection{Metallicity scale}
\label{metalcum}

Although the cluster metallicities adopted in G93 were chosen as the
most reliable estimates at the time, a revision can provide more
accurate values. The updated metallicities, together with Lick/IDS
values, are listed in Table~\ref{tab_zc}.

The [Fe/H] values for globular clusters stars that were adopted in the
Lick/IDS system, taken from several sources (see references in the
original papers), are basically in the Zinn \& West (1984, hereafter
ZW84) metallicity scale. This scale is based in a compilation of
metallicities from several parameters (mainly the photometric index
Q$_{39}$) tied to a high resolution scale using relatively old echelle
spectra. In the last years, and from high-quality high-dispersion
spectra and improved model atmospheres, several authors (Carretta \&
Gratton 1997, hereafter CG97; see also Rutledge, Hesser \& Stetson
1997) have derived a new homegeneous [Fe/H] scale which shows
significant deviations from the ZW84 scale. Although at this time no
consensus exists about which scale is more reliable, in this work we
have decided to adopt the new scale from CG97 for the following
reasons: (i) Rutledge et al. (1997) have shown that their near-IR Ca
index ($W^\prime$), the calibration of which is the objective of this
series of papers, correlates linearly with metallicities in the CG97
scale, but shows a non-linear behaviour with respect to the ZW84
values (in agreement with the previous work from CG97, which suggested
that the ZW84 may be non-linear with respect to the true [Fe/H] scale,
although note there is no a priori reason to expect a linear
behaviour); (ii) The main difference between the two scales occurs at
the high [Fe/H] end, where the ZW84 scale overestimates the
metallicities compared to the high-dispersion studies. The discrepancy
is specially important for M71, the highest [Fe/H] globular cluster of
our sample. We have checked that, if we assume for this cluster the
ZW84 [Fe/H] value and if we fit the strength of the near-IR Ca triplet
versus the atmospheric parameters (see Paper~III), we obtain
significant negative residuals in the sense that M71 should have a
lower [Fe/H] (and therefore closer to the CG97 value) if it were to
follow the behaviour of the rest of the sample stars (almost 700, most
of them from the field). Furthermore, in general, the globular cluster
residuals against the fitting functions derived in Paper~III are
significantly reduced when changing from the ZW84 to the CG97 scale.
We think that this adds further support to the reliability of the CG97
metallicity scale.

Concerning the open clusters stars, we have not introduced important
revisions to the adopted values in the Lick/IDS system. Basically, we
have chosen metallicities derived spectroscopically by E.D.~Friel and
collaborators (see Friel 1995). In all the cases, these are consistent
with more recent determinations. The main revision is for the cluster
NGC~7789, whose Lick/IDS [Fe/H] was taken from Twarog \& Tyson
(1985). The new metallicity of this cluster, listed by Friel (1995),
is fully consistent with the recent value derived from near-infrared
photometry by Vallenari, Carraro \& Richichi (2000).

\begin{table}
\caption{Adopted metallicities ([Fe/H]) for the cluster stars. Sources: (1)
Carretta \& Gratton (1997); (2) Rutledge, Hesser \& Stetson (1997), in the CG97 
scale; (3) Boesgaard \& Friel (1990); (4) Friel \& Boesgaard (1992);
(5) Friel \& Janes (1993); (6) Friel (1995).}
\label{tab_zc}
\begin{center}
\begin{tabular}{lccc}
\hline
Cluster  &  \multicolumn{1}{c}{Lick/IDS} & 
\multicolumn{1}{c}{This paper} & Source \\
\hline
M3      & $-1.70$ & $-1.34$ & 1 \\
M5      & $-1.30$ & $-1.11$ & 1 \\
M10     & $-1.50$ & $-1.41$ & 1 \\
M13     & $-1.50$ & $-1.39$ & 1 \\
M71     & $-0.56$ & $-0.70$ & 1 \\
M92     & $-2.20$ & $-2.16$ & 1 \\
NGC 6171 & $-0.99$ & $-0.95$ & 2 \\\\
Hyades   & $+0.13$ & $+0.13$ & 3 \\
Coma     & $-0.07$ & $-0.05$ & 4 \\
M67     & $-0.10$ & $-0.09$ & 5 \\
NGC 188  & $\ \  \,0.00$  & $-0.05$ & 6 \\
NGC 7789 & $-0.10$ & $-0.24$ & 6 \\
\hline
\end{tabular}
\end{center}
\end{table}

\subsection{Effective temperatures}

Due to the lack of direct measurements of effective temperatures for
cluster stars (mainly for globular clusters stars), we calculated
improved effective temperatures by means of colour-temperature
calibrations. Several recent papers have presented detailed
colour-temperature calibrations which take into account metallicity
and gravity effects. From a theoretical point of view, the work of
Houdashelt, Bell \& Sweigart \shortcite{HOU} (hereafter HBS) provides
a grid of colours which has been obtained from synthetic spectra and
put onto the observational system by comparing with field stars. It is
appropriate for stars having 4000 K $\leq$ {\it T}$_{\rm eff}$ $\leq$
6500 K, 0.0 $\leq$ log\,{\it g} $\leq$ 4.5 and --3.0 $\leq$ [Fe/H]
$\leq$ 0.0. On the other hand, the work of Alonso, Arribas \&
Mart\'{\i}nez-Roger
\shortcite{ALOdw} presents empirical colour calibrations for dwarfs and
subdwarfs with 4000 K $\leq$ {\it T}$_{\rm eff}$ $\leq$ 8000 K and
--2.5 $\leq$ [Fe/H] $\leq$ 0.0. An analogous work for giants having
3500 K $\leq$ {\it T}$_{\rm eff}$ $\leq$ 8000 K and --3.0 $\leq$
[Fe/H] $\leq$ 0.5 is presented in Alonso, Arribas \&
Mart\'{\i}nez-Roger \shortcite{ALOgi}. In the following, we will
denote the calibrations from Alonso et al. \shortcite{ALOdw}
and Alonso et al. \shortcite{ALOgi} as ALO.

\begin{figure}
\centerline{
\psfig{figure=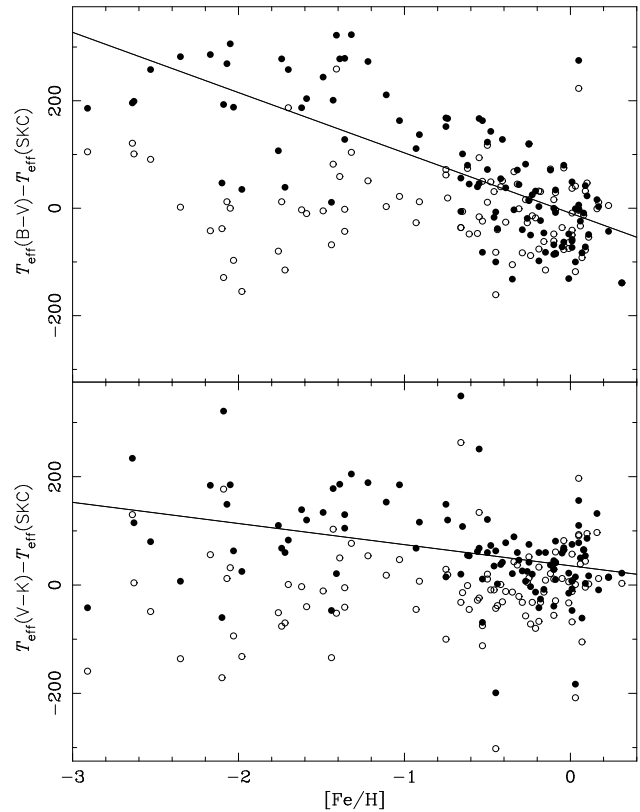}
}
\caption{Temperature differences between colour-temperature
calibrations ({\it T}$_{\rm eff}$($B-V$) and {\it T}$_{\rm eff}$($V-K$))
and the reference system ({\it T}$_{\rm eff}$(SKC)). Filled and open
circles are, respectively, data derived from the colour-{\it T}$_{\rm
eff}$ calibrations of HBS and ALO for a subsample of 103 library
stars. The solid line displays a least-squares fit to the filled
circles.}
\label{compHouAloSKC}
\end{figure}

It is important to keep in mind that our purpose is not only to derive
improved temperatures for the cluster stars but also to obtain an
homogeneous set of temperatures with the field stars. In order to
check whether the temperatures derived from the mentioned
colour-temperature relations are on our reference system, we selected
a subsample of 103 field stars from SKC which had both $B-V$ and $V-K$
data in the literature and compared their effective temperature with
the values predicted by the colour-temperature calibrations. Input
atmospheric parameters for the temperature predictions (log\,{\it g}
and [Fe/H]) are those published in SKC. $B-V$ input values were the
mean taken from the electronic database in Mermilliod, Mermilliod \&
Hauck \shortcite{MER}, while $V-K$ data were extracted, in order of
preference, from Johnson et al. \shortcite{John66}, Johnson, MacArthur
\& Mitchell \shortcite{John68}, Carney \shortcite{CAR83} and
Laird \shortcite{LAI}. The comparison is shown in
Fig.~\ref{compHouAloSKC} and reveals a systematic deviation of the
temperatures predicted by HBS for low metallicity stars, whereas no
significant deviations owing to metallicity effects are found in the
data from ALO. Thus, in order to preserve the homogeneity and quality
of the final data, we decided to use only ALO to derive the effective
temperatures of our library cluster stars. It is worth noting that, in
our stellar population synthesis model, these empirical
colour-temperature relations are preferred over the theoretical ones
(see Paper IV).

\begin{figure}
\centerline{
\psfig{figure=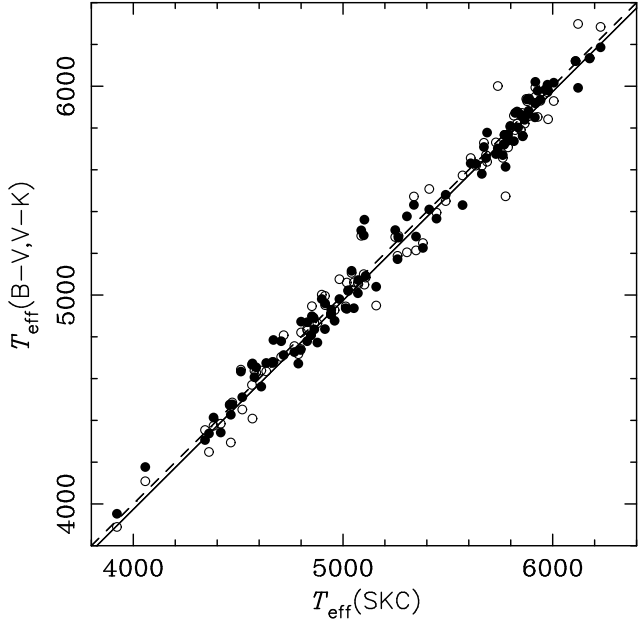}
}
\caption{Comparisons between effective temperatures derived
from ALO calibrations and those presented in SKC. Filled and open
circles, respectively, are stars with {\it T}$_{\rm eff}$ derived from
$B-V$ and $V-K$ calibrations. The dashed line is the one-to-one
relation. The solid line {\it T}$_{\rm eff}$ = {\it T}$_{\rm
eff}$(SKC) -- 26 marks a significant offset of 26 K for effective
temperatures derived from $B-V$. No significant deviation is observed
for effective temperatures predicted from $V-K$.}
\label{compALOSKC}
\end{figure}

Following the procedure described in Section\,\ref{calibracion}, we
have calibrated the effective temperatures predicted by ALO onto the
system of SKC (see Fig.~\ref{compALOSKC}). For temperatures derived
from the $B-V$ relations, a statistically significant offset of 26 K
was found, which has been applied to correct and bootstrap the
predicted data against the reference system. On the contrary, no
significant deviation has been found for temperatures derived from the
$V-K$ relations. In addition, the fact that the r.m.s standard
deviations from the fits are very similar ({\large $\sigma$}$_{\rm
B-V}$ = 75 K and {\large $\sigma$}$_{\rm V-K}$ = 78 K.) allows us to
deduce that the data quality of both predictions is the same.
Thus, final effective temperatures for cluster stars were calculated by
averaging the temperatures derived from $B-V$ and $V-K$ relations (when only
$B-V$ was available, final temperatures were exclusively derived from this
colour). Input metallicities are those previously established in Section
\ref{metalcum}, while gravities are the same as in G93 and W94. Concerning
the input colours, $V-K$ values where taken from G93, whilst $B-V$ colours are
from Mermilliod et al. \shortcite{MER} for Coma and Hyades, W94 for a few
horizontal--branch stars, and G93 for the rest of the sample. The reddening
corrections were applied using the color excesses given by G93.

It is important to note that, due to the validity range of the
colour-temperature calibrations, the effective temperatures of the hot
stars Hya vB 104, M5 II-53, M92 I-10, M92 II-23 and M92 XII-24 could
not be predicted in this way. In these cases, temperatures from W94
were kept.

\subsection{Surface gravities}

Surface gravities for most of the cluster stars were originally
derived by G93 by fitting the location of the stars in the HR diagrams
to different evolutionary tracks (see the original paper for details
on the tracks that were used). In the recent years, and thanks to the
new Hipparcos data, distances and absolute ages of most globular
clusters have been substantially modified (see e.g. Carretta et
al. 2000 and references therein). In this work we have studied what
changes should be introduced in the derived surface gravities to
account for the new, usually larger, distances and, therefore, younger
ages for the globular cluster stars. Changes in the effective
temperatures do not affect the derived gravities since the original
values were computed by matching only absolute magnitudes to avoid
uncertainties in the colour-$T_{\rm eff}$ determinations. Using the
relation $\Delta(\log g)=\Delta(\log (M/M_\odot)) + 0.4\Delta M_{\rm
bol}$ and applying the analytic formulae by Hurley, Pols \& Tout
(2000) to convert an age difference to a mass change for the stars in
the different evolutionary phases, we have checked that distance and
age effects tend to cancel each other, leading to systematic
differences in $\log g$ always below 0.05 dex (for all the globular
clusters with the exception of M71). This offset is below the
uncertainties associated with the employed evolutionary tracks and the
assumed metallicities and, therefore, we have decided not to change
the gravities derived in G93.

The case for M71 is more uncertain since the Hipparcos-based distance
modulus ($(m-M)_V=14.06$, Reid 1998) is rather larger than that
employed in G93 ($(m-M)_V=13.40$). Furthermore, Salaris \& Weiss
(1998) have lowered the colour-magnitude derived age from the 18 Gyr
assumed in G93 to only 9.2 Gyr. We must note here that if we assume
the distance modulus from Reid (1998), we are unable to fit the
location of the stars in the M71 horizontal branch to theoretical
isochrones, while a modulus around $(m-M)_V=13.60$ (employed in the
colour-magnitude diagram analyses of Hodder et al. 1992, Geffert \&
Maintz 2000, and Rosenberg et al. 2000) can explain the position of
these stars. Assuming the latter distance, we have checked that the
original gravities should be decreased or increased by $\sim 0.05$ dex
for ages of 9.2 and 15 Gyr respectively. We have therefore assumed
that the gravities given by G93 are correct within the uncertainties
associated with the absolute parameters of this cluster.

To summarize, we have not introduced any change to the gravities given
by G93. Systematic errors of the order of 0.10 dex (well below the
random errors given in Table~\ref{errores} for the field stars) may
still exists for the absolute gravities of the cluster stars. The
detailed investigation of these errors is beyond the scope of this
work, and, in any case, they do not represent a major drawback for the
purposes of this series of papers.

\begin{figure*}
\centerline{
\psfig{figure=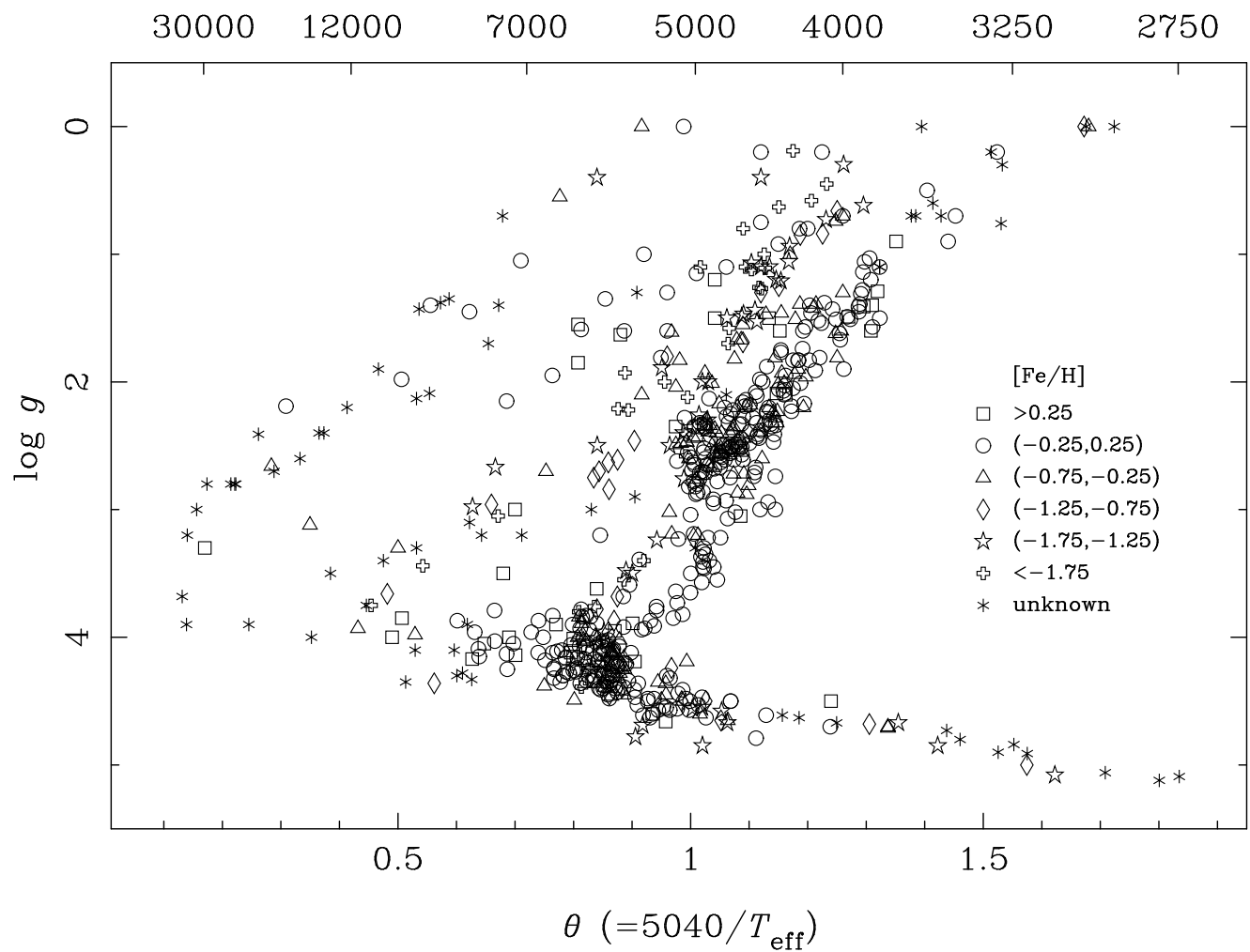}
} 
\caption{Gravity--temperature diagram for the library
stars. Different symbols are used to indicate stars of different
metallicities, as shown in the key. The upper scale gives effective
temperatures in K}
\label{diagHR}
\end{figure*}

\section{Summary}

The uncertainties in the input atmospheric parameters are one of the
main sources of potential errors when computing the predicted
line-strengths of composite stellar systems using evolutionary
synthesis models. In this paper we have derived a reliable, and highly
homogeneous, set of atmospheric parameters (2747 K $<$ {\it T}$_{\rm
eff}$ $<$ 38367, 0.00 $<$ log\,{\it g} $<$ 5.12 and --3.45 $<$ [Fe/H]
$<$ +0.60) for the 706 stars which constitute a new stellar library in
the near-infrared spectral range of the calcium triplet ($\lambda$
8350--9020 \AA). Systematic deviations between parameters from
different sources have been calibrated and corrected by bootstrapping
them onto a reference system. Fig.~\ref{diagHR} shows the complete
stellar library in the parameter space of temperature and gravity for
various metallicity-ranges. In the forthcoming papers, the results of
this work will be used to derive empirical fitting functions for the
calcium triplet index (Cenarro et al. 2001b, Paper~III) and to predict
the integrated spectral energy distributions of stellar populations in
the near-IR spectral range (Vazdekis et al. 2001, Paper~IV).
Moreover, the utility of the new set of improved parameters goes
beyond the objectives of this series. In particular, it should
represent a basic ingredient for the new generation of spectral
synthesis work and to improve the existing empirical calibrations of
other relevant spectral features.

\section*{ACKNOWLEDGMENTS}

We are indebted to the anonymous referee for very useful suggestions.
Work on writing this paper was partially developed while AC was a
visitor at the Department of Physics (University of Durham) and at the
School of Physics and Astronomy (University of Nottingham), whose
hospitality is acknowledged gratefully. This research has made use of
the Simbad database (operated at CDS, Strasbourg, France), the NASA's
Astrophysics Data System Article Service, and the Hipparcos Input
Catalog. AV acknowledges the support of the PPARC rolling grant
'Extragalactic Astronomy and Cosmology in Durham 1998-2002'. This work
was supported in part by a British Council grant within the
British/Spanish Joint Research Programme (Acciones Integradas). AC
acknowledges the Comunidad de Madrid for a Formaci\'on de Personal
Investigador fellowship.  This work was supported by the Spanish
Programa Sectorial de Promoci\'on del Conocimiento under grants
No. PB96-610 and AYA2000-977

\label{lastpage}
\end{document}